\documentclass[useAMS,usenatbib]{mn2e}
\usepackage{graphicx}
\usepackage{epstopdf}
\usepackage{amssymb}
\usepackage{amsfonts}

\def\mnras{MNRAS}
\def\nat{Nature}
\def\apj{ApJ}
\def\apjl{ApJL}
\def\aap{A\&A}
\def\prd{Phys. Rev. D}
\def\nar{New A. Rev.}
\def\na{New A.}
\def\araa{ARA\&A}
\def\jcap{J. Cosmology Astropart.Phys.}
\def\ssr{SSRv}

\title[Radio Emission in the Cosmic Web]{Radio Emission in the Cosmic Web}
\author[P. A. Araya-Melo et al.]{\parbox{\textwidth}{Pablo A. 
Araya-Melo$^{1}$\thanks{Currently at Georges Lema\^{i}tre Centre for Earth and 
Climate Research, Universit\'{e} catholique de Louvain, Louvain-la-Neuve, 
Belgium, pablo.arayamelo@uclouvain.be}, Miguel A. Arag\'{o}n-Calvo$^2$, Marcus 
Br\"{u}ggen$^{1}$ \& Matthias Hoeft$^3$}
\vspace*{4pt}\\
$^1$Jacobs University Bremen, Campus Ring 1, 28759 Bremen, Germany\\
$^2$The Johns Hopkins University, 3701 San Martin Drive, Baltimore, MD 21218, 
USA\\
$^3$Th\"{u}ringer Landessternwarte, Sternwarte 5, 07778 Tautenburg, Germany}
\begin{document}

\date{}

\pagerange{\pageref{firstpage}--\pageref{lastpage}} \pubyear{2012}

\maketitle

\label{firstpage}

\begin{abstract}
We explore the possibility of detecting radio emission in the \emph{cosmic web}
by analyzing shock waves in the MareNostrum cosmological simulation. This 
requires a careful calibration of shock finding algorithms in Smoothed-Particle
Hydrodynamics simulations, which we present here. Moreover, we identify the 
elements of the cosmic web, namely voids, walls, filaments and clusters with 
the use of the SpineWeb technique, a procedure that classifies the structure 
in terms of its topology. Thus, we are able to study the Mach number 
distribution as a function of its environment. We find that the median Mach 
number, for clusters is 
$\mathcal{M}_{\mathrm{clusters}}\approx1.8$, for filaments is 
$\mathcal{M}_{\mathrm{filaments}}\approx 6.2$, for walls is 
$\mathcal{M}_{\mathrm{walls}}\approx 7.5$, and for voids is 
$\mathcal{M}_{\mathrm{voids}}\approx 18$. We then estimate the radio emission 
in the cosmic web using the formalism derived in Hoeft \& Br\"{u}ggen (2007).
We also find that in order to match our simulations with observational data
(e.g., NVSS radio relic luminosity function), a fraction of energy dissipated 
at the shock of $\xi_{\mathrm{e}}=0.0005$ is needed, in contrast with the 
$\xi_{\mathrm{e}}=0.005$ proposed by Hoeft et al. (2008). We find that 41\% of 
clusters with $M \ge 10^{14} M_{\odot}$ host diffuse radio emission in the form 
of radio relics. Moreover, we predict that the radio flux from filaments should
be $S_{150 MHz}\sim 0.12$ $\mu$Jy at a frequency of 150 MHz.
\end{abstract}


\section{Introduction}

In the standard theory of formation history, matter evolved from small 
perturbations in the primordial density field into a complex structure of sheets
and filaments with galaxy clusters at the intersections of this filamentary 
structure. Galaxy surveys, such as the 2dF-GRS  \citep[2 degree Field Galaxy 
Redshift Survey, ][]{colless2003} and the SDSS \citep[Sloan Digital Sky 
Survey, ][]{tegmark2004}, have revealed a complex network of filamentary 
nature, which has become necessary known as the cosmic web \citep{bond1996}. 

In the evolution of the cosmic web, baryons in the diffuse intergalactic medium
accelerate towards dark matter halos under the growing influence of gravity and
go through shocks that heat them. These cosmological shocks are a ubiquitous 
consequence of cosmic structure formation. They are tracers of the large-scale 
structure and contain information about its dynamical history. Gas in walls and
filaments follows the gravitational potential towards clusters of galaxies, 
colliding with the intracluster medium (ICM) at speeds of 
$\gtrsim 1000$ km s$^{-1}$. While cosmological shocks heat the ICM to 
temperatures of $\sim 10^{7}-10^{8}$ K, filaments are heated to temperatures of 
$10^{5}-10^{7}$ K, making their detection at present time challenging, since at 
this temperature there are hardly any emission lines. Therefore, the study of 
shocks may provide an independent and complementary way to study the Warm-Hot 
Intergalactic Medium (WHIM), the low density intergalactic medium of the cosmic
web that is believed to host the majority of the fraction of the baryon density.

Some cosmological shocks are associated with diffuse radio emission caused by
electrons accelerated to relativistic velocities by Fermi acceleration 
\citep{ensslin1998, roettiger1999, miniati2001}. This diffuse radio 
emission, without galaxy counterpart, is usually divided into two classes, 
namely \emph{radio halos} and \emph{radio relics}. Radio halos are unpolarized 
and have diffuse morphologies that are similar to those of the thermal X-ray 
emission of the cluster gas \citep{giovannini2006}. They are usually found in 
the center of clusters with significant substructure 
\citep[see e.g.,][]{cassano2010}.

Radio relics, on the other hand, are typically located near the periphery of 
the cluster. They have been observed in several merging galaxy clusters. They 
often exhibit sharp emission edges and many of them show strong radio 
polarization. Once accelerated, the electrons are short-lived because of the 
inverse Compton scattering and synchrotron energy losses, and their spectrum 
rapidly steepens from the shock edge \citep{giacintucci2008, vanweeren2009}
The synchrotron nature of this radio emission indicates the presence of cluster
magnetic fields of the order of $\sim 0.1-1$ $\mu$G \citep{feretti2008}. 

Cosmic shocks in large-scale structures have been investigated in a number of
semi-analytical \citep{gabici2003, berrington2003, keshet2003, meli2006} and 
numerical work \citep{miniati2000, miniati2001, miniati2002, ryu2003, kang2005,
pfrommer2006, pfrommer2007, pfrommer2008,jubelgas2008, hoeft2008, skillman2008,
battaglia2009, vazza2009, skillman2010}. Some studies have focused on the 
non-thermal emission from galaxy clusters by modeling discretized cosmic ray 
(CR) energy spectra. A series of papers explored the dynamical impact of cosmic
ray (CR) protons on hydrodynamics in cosmological SPH simulations 
\citep{pfrommer2006, ensslin2007, jubelgas2008}. Observationally, detecting 
shocks waves in large-scale structures is still challenging, since they usually
develop in the external regions of galaxy clusters, where the X-ray emission is
faint. However, a few merger shocks have been detected in nearby X-ray bright 
galaxy clusters \citep{markevitch2005, markevitch2006, solovyeva2008} and may 
be possible associated with a single or double radio relics discovered in a 
number of galaxy clusters \citep[e.g.,][]{roettgering1997, markevitch2005, 
bagchi2006, bonafede2009, giacintucci2008}. 

In numerical studies, \cite{miniati2000} studied the quantitative properties 
of large-scale shocks produced by gas during the formation of cosmic structures
by means of hydrodynamical simulations. They showed that shocks form abundantly
in the course of structure formation and their topology is very complex and 
highly connected. They also stated that considering the large size and long 
lifetime of shocks, they are potentially interesting sites for cosmic-ray 
acceleration. \cite{ryu2003} (see also \citealt{kang2005}) found that shocks 
form around sheets, filaments and knots of mass distribution when the gas in 
void regions accretes onto them.

\cite{pfrommer2006} studied the properties of structure formation shock waves in
cosmological SPH simulations, allowing them to study their role in the 
thermalization of the plasma as well as for the acceleration of relativistic
CRs through diffusive shock acceleration. They find that most of the energy
is dissipated in weak internal shocks, with Mach number $\mathcal{M}\sim 2$. 
On the other hand, collapsed cosmological structures are surrounded by external
shocks with much higher Mach numbers ($\mathcal{M}\sim 100$), but they play 
only a minor role in the energy balance of thermalization. \cite{skillman2008} 
computed the production of CRs in a cosmological simulation volume, finding 
that CRs are dynamically important in galaxy clusters. They also found that 
shocks with low Mach number typically trace mergers and complex flows, while 
those with mild and high Mach number ($\mathcal{M}> 5$) generally follow 
accretion onto filaments. In a similar study, \cite{vazza2009} analyzed the 
properties of large-scale shocks. They find that the bulk of the energy in 
galaxy clusters is dissipated at weak shocks, with Mach numbers 
$\mathcal{M}\approx 1.5$, although slightly stronger shocks are found in the
external regions of merging clusters.

\cite{pfrommer2008} used GADGET simulations of a sample of galaxy
clusters, implementing a formalism for CR physics on top of radiative 
hydrodynamics. They modeled relativistic electrons that are accelerated at
cosmic formation shocks and produced in hadronic interactions of CRs with
protons of the ICM. They found that the radio emission in clusters is dominated
by secondary electrons. Only at the location of strong shocks the contribution 
of primary electrons may dominate. Later, \cite{battaglia2009} studied the 
radio emission for a sample of 10 galaxy clusters extracted from zoomed 
cosmological simulations. They determined the radio luminosity, spectral index 
and Faraday rotation measure for regions of the relics and concluded that 
upcoming radio telescopes, such as LOFAR, MWA, LWA and SKA will discover a 
substantially larger sample of radio relics. SKA should probe the macroscopic 
parameters of plasma physics in clusters.

\cite{hoeft2008} investigated the diffuse radio emission from clusters in 
cosmological SPH simulations. They found that the maximum diffuse radio 
emission in clusters depends strongly on their X-ray temperature. They also 
found that the so-called accretion shocks cause only very little radio 
emission. They conclude that a moderate efficiency of shock acceleration, 
namely $\xi_{\mathrm{e}}=0.005$, together with moderate magnetic field, namely 
0.07-0.8 $\mu$G, in the region of radio relic are sufficient to reproduce the 
number density and luminosity of radio relics. \cite{skillman2010}, using AMR 
simulations, also studied radio emission in galaxy clusters. They investigated 
scaling relations between cluster parameters such as synchrotron power, mass 
and X-ray luminosity.

In this paper we follow the approach of \cite{hoeft2008} but, instead of
focusing on galaxy clusters, we investigate whether it is possible to detect
radio emission in the entire cosmic web. Due to the low temperature of the 
accreting gas, the Mach number of external shocks - filaments - is high, 
extending up to $\mathcal{M}\sim 100$ or higher \citep{ryu2003}. Therefore, the
total radio power generated by infall along filaments will be significantly 
weaker than infall into a cluster. We will try to determine this total radio
power. To this end, we use the approach for detecting and identifying shock 
fronts and their characteristic Mach number developed by \cite{hoeft2008}. This
shock finder is applied to a large $N$-body/SPH simulation. For computing the 
radio emission, we follow the model elaborated in \cite{hoeft2007}. They 
assumed that electrons are accelerated by diffuse shock acceleration and cool 
subsequently by synchrotron and inverse Compton losses. Consequently, the radio
emission can be expressed as a function  of downstream plasma properties, Mach 
number and surface area of the shock front. In order to correctly assign a 
surface area to the SPH particles, we calibrate the radio emission using shock 
tubes experiments. Of primary importance is the proper identification of the 
different structures that reside in the cosmic web. To this end, we apply the 
SpineWeb technique (\citealt{aragoncalvo2010a}, see also 
\citealt{aragoncalvo2010c}) to the entire simulation. The SpineWeb procedure 
correctly identifies voids, walls, filaments and clusters present in the cosmic
web. It is a powerful method that deals with the topology of the underlying 
density field of the cosmic web. This allows us to study shock properties and 
radio emission in the different structures that constitutes the cosmic web. 
This goes beyond the traditionally internal versus external shock 
classification suggested by \cite{ryu2003} and the method of characterizing 
shocks by their pre-shock overdensity and temperature proposed by 
\cite{skillman2008}. Applying the radio emission model to the simulation leads 
to radio-loud shock fronts which we are able to locate within the cosmic web.

This paper is organized as follows. In Section~\ref{sec:computer_simulation} we
summarize the characteristics of the $N$-body/SPH simulation. 
Section~\ref{sec:spineweb} describes the Spineweb method. In 
Section~\ref{sec:hydroshocks} we give a brief description of the physics 
involving non-radiative shocks and depict how we find shocks in SPH 
simulations. In order to calibrate the radio emission model, we use shock tube 
tests which are described in Section~\ref{sec:shocktubes}. The radio emission 
model is sensitive to the shock surface area. Using the results of the shock 
tubes, in Section~\ref{sec:shockarea} we described the method used in order to 
precisely determine the area of the shock front. In Section~\ref{sec:results} 
we present our results for the shock fronts and radio emission in the 
MareNostrum Universe. In Section~\ref{sec:conclusions} we summarize our 
findings.

\section{The Computer Simulation}
\label{sec:computer_simulation}

In order to study cosmological shock waves in the large-scale structure of the
Universe, we use the \emph{MareNostrum Universe} simulation 
\citep{gottloeber2007}, one of the largest cosmological simulations available. 
It assumes a standard flat $\Lambda$CDM Universe with cosmological parameters 
$\Omega_{m,0}=0.3$, $\Omega_{b,0}=0.045$, $\Omega_{\Lambda,0}=0.7$ and $h=0.7$, 
where the Hubble parameter is given by  $H_{0}=100h$ km s$^{-1}$Mpc$^{-1}$. The 
normalization of the power spectrum is $\sigma_{8}=0.9$. The simulation box has
a side length of 500$h^{-1}$Mpc and contains 1024$^{3}$ gas and 1024$^{3}$ dark 
matter particles, with masses of 
$m_{\mathrm{gas}}=1.45\times10^{9}h^{-1}$M$_{\odot}$ and 
$m_{\mathrm{dm}}=8.24\times10^{9}h^{-1}$M$_{\odot}$. The initial conditions are 
followed from a redshift of $z=40$ until the present time ($z=0$) using the 
massively parallel tree $N$-Body/smoothed particle hydrodynamics (SPH) code 
GADGET-2 \citep{springel2005}. The results presented here are at present time
($z=0$). Radiative processes or star formation are not included in the 
simulation. The Plummer-equivalent softening was set at 
$\epsilon_{Pl}=15$ $h^{-1}$kpc in comoving units, and the SPH smoothing length 
was set to the distance to the 40th nearest neighbor of each SPH particle. 
Smoothing scales are not allowed to be smaller than the gravitational softening
of the gas particles.

To extract the groups and galaxy clusters present in the simulation, we use HOP
\citep{eisenstein1998}. HOP assigns a density to every particle by smoothing
the density field with a spline cubic kernel using the $n$ nearest neighbors of
a given particle. Particles are then linked by associating each particle to the
densest particle from the list of its closest neighbor. This process is repeated
until it reaches a particle that is its own highest density neighbor. All 
particles linked to a local density maximum are identified as a group. At this
stage, no distinction between a high-density region and its surrounding has been
make. To identify halos above a certain density threshold, a regrouping merging
procedure is performed. The code first includes only particles that are above
some density threshold. Then, it merges all groups for which the boundary
density between them exceeds a certain density value. Finally, all groups 
identified must have one particle that exceeds a density peak to be accepted as
an independent group. We associate this density peak with the virial density
needed for a spherical region to be in virial equilibrium, 
$\Delta_{\mathrm{vir}}$. The value of $\Delta_{\mathrm{vir}}$ is obtained from the 
solution to the collapse of a spherical top-hat perturbation under the 
assumption that the object has just virialized. For the cosmology described 
here, at present time its value is $\Delta_{\mathrm{vir}}\approx 337$. 

In using HOP, we use the dark matter particles of the MareNostrum simulation. 
Once we identified the virialized halos of the sample, we take the position of
the densest dark matter particle as a first estimate of the center of mass. 
Subsequently, we add the gas particles, grow a sphere around the center of mass
and begin iterating, shrinking the sphere around the (new) center of mass 
until we reach a minimum of 50 particles. This ensures a correct 
identification of the center of mass taking into account gas and dark matter 
particles. Once we have the final center of mass, we grow a sphere around it 
that encloses a value of $\Delta_{\mathrm{vir}}\approx 337$. Given that we want to
study radio emission in the cosmic web, the nodes of the filaments will be 
consider as galaxy clusters with masses $M>10^{14}h^{-1}$M$_{\sun}$. We find that
the MareNostrum Universe contains 3865 clusters of galaxies, the most massive 
with a mass of $M=2.5\times 10^{15}h^{-1}$M$_{\sun}$ and virial radius 
$R_{\mathrm{vir}}=2.75h^{-1}$Mpc.

\section{The SpineWeb procedure}
\label{sec:spineweb}

The characterization of the Large-Scale Structure was done using the SpineWeb
method (for a thorough explanation see \citealt{aragoncalvo2010a}, see also
\citealt{aragoncalvo2010c}). This method classifies the Cosmic Web into four 
basic morphological and dynamical components: voids, walls, filaments and 
clusters. The SpineWeb method is based on the Watershed transform 
\citep{beucher1979} and its cosmological implementation: the Watershed Void 
Finder (WVF) \citep{platen2007}. The idea behind the WVF is to segment
the density field into individual basins by "flooding" it (in analogy to a
landscape). The SpineWeb method takes this analogy and
goes one step further by directly relating the topology of the density
field and the boundaries between basins by observing the
simple relation between number of adjacent basins and topology: walls
correspond to the boundary between two voids. Filaments+clusters are found
at the intersection of three or more voids, which translates to the
intersection of three or more walls: 

\begin{equation}
\mathcal{N}_{\textrm{\tiny{voids}}}\qquad \left\{
 \begin{array}{rrl}
     \quad =  \;\;\; 1, \qquad & \mathrm{void} \\
     \quad =  \;\;\; 2, \qquad & \mathrm{wall} \\
     \quad \geq \;\;\; 3, \qquad & \mathrm{spine} \\
     \quad \qquad & \mathrm{(filament + clusters)} \\
 \end{array} \right.
\label{eq:neighbour_morphology}
\end{equation}

The resulting morphological characterization is parameter-free. The
analysis presented here is also based on the most recent
implementation of the SpineWeb method which extends the original formalism
on a multiscale hierarchical fashion making it also scale-independent.

In practice we start our analysis by generating a low resolution version
of the original simulation using the averaging procedure described in 
\cite{klypin2001}. The lower particle resolution was chosen to correspond to a 
inter-particle separation of  approx. 3$h^{-1}$Mpc at the initial conditions. 
With this ``linear-regime'' filtering we avoid small-scale substructure at the 
present time while keeping the characteristic large-scale anisotropy of the 
cosmic web. From the particle distribution we compute the density field using 
the DTFE method \citep{schaap2000,schaap2007,vdw2009}. The field is sampled at 
27 different points inside each voxel and its mean is used to estimate the 
density for that particular voxel. From the density field we compute the 
watershed transform and identify the boundaries of the basins (i.e. the 
watershed transform itself). Finally for each voxel in the watershed transform 
we apply the criteria outlined in Eqn.~\ref{eq:neighbour_morphology}. The 
complete procedure takes a few minutes on a regular workstation.

\begin{figure}
\centering
\includegraphics[width=0.50\textwidth]{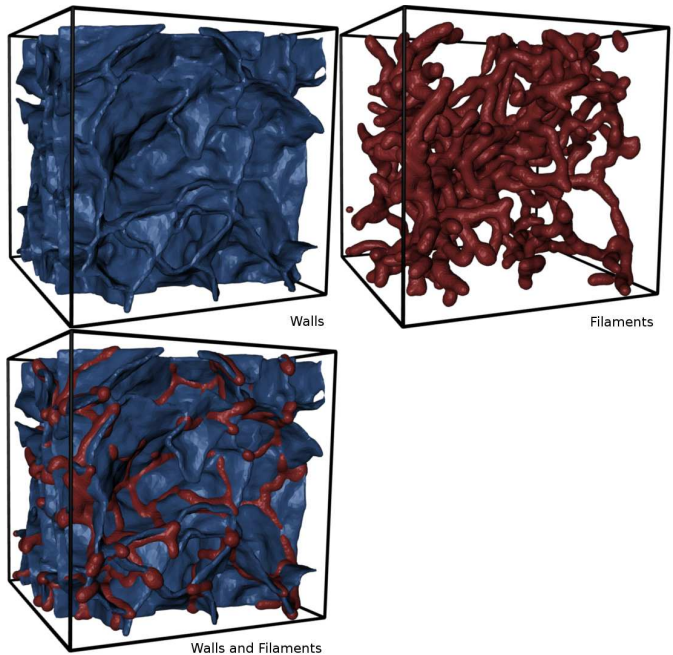}
\caption{\small{An example of how the SpineWeb procedure divides the cosmic 
web. Shown are surfaces enclosing voxels which are identified as belonging to 
walls (blue, top left frame) and filaments (red, top right frame) within a 
cubic region of 50$h^{-1}$Mpc. The bottom frame shows the same region with both
components connected and intertwined.}}
\label{fig:spine}
\end{figure}

The SpineWeb method provides a complete framework for the characterization 
of the cosmic web into its basic morphological constituents: voids, walls,
filaments and clusters. It has no free parameters and being fundamentally a 
topological measure it is highly robust against noise in the density field.
Fig.~\ref{fig:spine} shows an example of how the SpineWeb correctly identifies 
the different elements of the cosmic web. Shown is a full 3D network of 
filaments (red) and walls (blue). It is clearly visible the three-dimensional 
nature of the filament-wall network. Filaments define an interconnected web and 
walls fill the spaces in between the filaments forming a closed "watertight" 
network of voids.

\section{Hydrodynamical Shocks and Shock Finder}
\label{sec:hydroshocks}

In the formation of structures in the Universe, the bulk velocity of the gas
flow often exceeds the local sound speed. As a result, the shock surface 
separates two regions: the \emph{upstream regime} or pre-shock regime, and the
\emph{downstream regime} or post-shock regime. The plasma in the upstream 
regime moves with a velocity $v_{\mathrm{u}}$ towards the shock front, while the 
downstream plasma departs with velocity $v_{\mathrm{d}}$. Most of the incident 
kinetic energy flux is converted into thermal energy. As the plasma passes 
through the shock front, mass, momentum and energy fluxes are conserved, which 
is expressed in the Rankine-Hugoniot relations:

\begin{eqnarray}
\rho_{\mathrm{u}}v_{\mathrm{u}}\,&=&\,\rho_{\mathrm{d}}v_{\mathrm{d}}\,,\nonumber\\
P_{\mathrm{u}}+\rho_{\mathrm{u}}v_{\mathrm{u}}^{2}\,&=&
\,P_{\mathrm{d}}+\rho_{\mathrm{d}}v_{\mathrm{d}}^{2}\,,\nonumber\\
\frac{v_{\mathrm{u}}^{2}}{2}+u_{\mathrm{u}}+\frac{P_{\mathrm{u}}}{\rho_{\mathrm{u}}}\,
&=&\,\frac{v_{\mathrm{d}}^{2}}{2}+u_{\mathrm{d}}+\frac{P_{\mathrm{d}}}{\rho_{\mathrm{d}}}\,,
\label{eq:rankine-hugoniot}
\end{eqnarray}
where $\rho$ denotes the mass density, $P$ is the pressure and $u$ is the 
specific internal energy, and the velocities are measured in the rest-frame of 
the shock surface. The strength of a shock is given by the Mach numbers 
$\mathcal{M}$,
\begin{equation}
\mathcal{M}\,=\,\frac{v_{\mathrm{u}}}{c_{\mathrm{u}}}\,,
\label{eq:machnumber}
\end{equation}
where $c_{s}$ denotes the upstream sound speed which depends on the the specific
internal energy by $c_{\mathrm{u}}=\sqrt{\gamma(\gamma-1)u_{\mathrm{u}}}$. 
Combining Eqns.~\ref{eq:rankine-hugoniot} and \ref{eq:machnumber}, and assuming
that the plasma obeys the polytropic relation, we can write for the Mach number
\begin{equation}
\mathcal{M}^{2}\,=\,\frac{r}{\gamma}\frac{qr^{\gamma}-1}{r-1}\,,
\label{eq:machnumber2}
\end{equation}
where $r$ and $q$ are the compression and entropy ratios. The difference 
between downstream and upstream velocity can be written as
\begin{equation}
v_{\mathrm{d}}-v_{\mathrm{u}}\,=\,\mathcal{M}\frac{r-1}{r}u_{\mathrm{u}}\,.
\label{eq:updownvel}
\end{equation}

The Mach number estimator, which will be explained in the next section, 
computes the entropy ratio and the ratio 
$(v_{\mathrm{d}}-v_{\mathrm{u}})/u_{\mathrm{d}}$ for each SPH particle in the 
simulation. In order to obtain the Mach number, we tabulate the relations 
between $q$, $r$, $\mathcal{M}$ and $\mathcal{M}(r-1)/r$, allowing us to simple
read the Mach number from a table.

\subsection{Shock finder and Mach number estimate}
\label{sec:shockfinder}

In order to calculate the radio emission present in the MareNostrum simulation,
the first step is to identify shock fronts, and from this, derive the Mach 
number, downstream (postshock) temperature and electron density. The method we
employ is the one proposed and used by \cite{hoeft2008}. For a detailed 
description of the method, we refer to the mentioned paper. Here, we will give
the basic details.

The first step consists of computing the entropy gradient, $\nabla S$, for each
SPH particle. The entropy gradient gives the direction of the shock normal 
pointing into the downstream direction. We define an associated upstream and
downstream position,
\begin{equation}
\mathbf{x_{li}}\,=\,\mathbf{x_{i}}+f_{h}h_{i}\mathbf{n}^{1}_{i}\,,
\label{eq:updownposition}
\end{equation}
where $l=u,d$ denotes upstream or downstream, with the downstream position in 
the opposite direction, $\mathbf{x}_{i}$ is the position of the SPH particle 
$i$, $h_{i}$ is the smoothing length of particle $i$ and $\mathbf{n}^{1}_{i}$
denotes the shock normal, $\mathbf{n}^{1}_{i}=-\nabla S/|\nabla S|$.

The upstream and downstream velocity are computed, together with the internal
energy and density, using the usual SPH scheme. The upstream velocity is given
by $v_{ui}+v_{sh}=\mathbf{v}(\mathbf{x}_{ui})\cdot\mathbf{n}^{1}_{i}$, and the same
for the downstream velocity, but using the downstream position. $v_{sh}$ is the 
velocity of the shock front in the rest-frame of the simulation. It is important
to notice that the velocity field can also show perpendicular components. We 
compute these components in a similar way to the upstream and downstream 
velocities, $v_{i}^{k\pm}=
\mathbf{v}(\mathbf{x}_{i}\pm f_{h}h_{i}\mathbf{n}^{k}_{i})\cdot\mathbf{n}^{k}_{i}$,
where the three vectors $\mathbf{n}^{1}$, $\mathbf{n}^{2}$ and $\mathbf{n}^{3}$
form an orthonormal base.

Once this is done, for those particles that belong to a shock front we demand 
that the velocity has to be divergent in the direction of the shock normal, 
i.e., $(v_{\mathrm{d}}-v_{\mathrm{u}})>0$. Also, the velocity difference in the 
directions perpendicular to the shock normal have to be smaller than that 
parallel to the shock, i.e., 
$|v_{k}^{+}-v_{k}^{-}|<(v_{\mathrm{d}}-v_{\mathrm{u}})/2$. 

Finally, to compute the Mach number, there are two options: it is possible to
use the entropy ratio $q=S_{d}\S_{u}$ ($S$ is the entropy) or use the ratio
$(v_{\mathrm{d}}-v_{\mathrm{u}})/c_{\mathrm{u}}$, where $c_{\mathrm{u}}$ is the sound 
speed in the upstream region (see Section~\ref{sec:hydroshocks}). In order to 
have a conservative estimate for the Mach number, the smaller value of these two
options is used.

\section{Radio Emission Model}
\label{sec:radioemission}

In this section, we will briefly describe the radio emission model. For a 
complete description, we refer to \cite{hoeft2007} and \cite{hoeft2008}.

The first assumption of the model is that radio emission is produced by 
electrons accelerated to ultra-relativistic speeds required for synchrotron and
inverse Compton radiation. These electrons are accelerated to a power-law
distribution that is related to the Mach number from diffusive shock 
acceleration theory. As for the magnetic field
in the downstream region, it is assumed that, on average, simply obeys
flux conservation to follow: 

\begin{equation}
\frac{B_{\mathrm{d}}}{B_{\mathrm{ref}}}\,=\,\left(\frac{n_{\mathrm{d}}}{10^{-4}
\textrm{cm}^{-3}}\right)^{2/3}\,,
\label{eq:magfield}
\end{equation}
where $B_{\mathrm{ref}}$ is $0.1\mu G$ and $n_d$ is the number density 
in the downstream region. This is a simple assumption since a detailed
model for the generation and evolution of magnetic fields is complex and beyond
the scope of this paper (see, e.g., \citealt{bonafede2010} for a recent study 
on magnetic fields in the Coma cluster). 

There are several indications that the magnetic ﬁeld intensity 
decreases going from the center to the periphery of a cluster. The exact power 
with which it varies with density is unclear, however. Some simulations suggest
a higher power than that given by flux conservation which is attributed to the 
work of a fluctuation dynamo (see e.g. \cite{dolag2008} and 
\cite{brueggen2005}). Faraday rotation measurements in the Coma cluster 
\cite{bonafede2010} give a best fit for $B\propto n^{0.5}$. Such a power would 
decrease the luminosity of cluster shocks by about 20\%. Outside of clusters 
there are even fewer observational constraints. Recent constraints from TeV 
$\gamma$-ray sources have placed lower limits on extragalactic magnetic fields 
(EGMFs). The charged particles from pair cascades are deflected by EGMFs, 
thereby reducing the observed point-like flux. Dolag et al. (2011) have 
calculated the fluence of 1ES 0229+200 as seen by Fermi-LAT for different EGMF.
The non-observation of 1ES 0229+200 by Fermi-LAT suggests that the EGMF fills 
at least 60\% of space with fields stronger than $10^{-15}$ G assuming that the
source is stable for at least $10^4$ yr. The fields outside clusters, however, 
will have a negligible effect on the radio luminosity function since the by far
the largest part of the radio luminosity originates from shocks in or near 
clusters.

The radio emissivity, $P(e,\nu_{\mathrm{obs}})$, which depends on three 
quantities: i) the energy of the electron energy ($e$), ii) the observing 
frequency ($\nu_{\mathrm{obs}}$), and iii) the magnetic field strength ($B$), is
then given by:

\begin{eqnarray}
\frac{\mathrm{d}P(\nu_{\mathrm{obs}})}{\mathrm{d}\nu}\,=\,
6.4\times10^{34}\mathrm{erg}\,\textrm{s$^{-1}$}\textrm{Hz$^{-1}$}\,
\frac{A}{\textrm{Mpc$^{2}$}}\frac{n_{e}}{10^{-4}}\textrm{cm$^{-3}$}\nonumber\\
\times\frac{\xi_{\mathrm{e}}}{0.05}\left(
\frac{\nu_{\mathrm{obs}}}{1.4\mathrm{GHz}}\right)^{-s/2}\times\left(
\frac{T_{\mathrm{d}}}{7\mathrm{keV}}\right)^{3/2}\nonumber\\
\times\frac{(B/\mu G)^{1+s/2}}{(B_{\mathrm{CMB}}/\mu G)^{2}+(B/\mu G)^{2}}
\Psi(\mathcal{M},T_{\mathrm{d}})\,,
\label{eq:total_emission2}
\end{eqnarray}
where $n_{e}$ is the electron density, $\xi_{\mathrm{e}}$ is the 
fraction of energy dissipated by the electron at the shock front and $s$ is the
compression ratio at the shock front. The radiation spectral index is related 
to the electron spectral index by $\alpha=(s-1)/2$. $B_{\mathrm{CMB}}$ has a 
value of $3.24=\mu G (1+z)^{2}$. This magnetic field is defined as the magnetic
field corresponding to the energy density of the CMB.

As stated before, in order to calculate the radio emission, we need the Mach 
number, $\mathcal{M}$, the downstream temperature, $T_{\mathrm{d}}$, the electron
density, $n_{e}$, the magnetic field, $B$, and the shock surface area, $A$. The
shock finder described in Sect.~\ref{sec:shockfinder}, besides locating shock 
discontinuities in a SPH simulation, also provides estimates for the Mach 
number and for the shock normal. This normal vector gives a position which is 
sufficiently downstream to determine $T_{\mathrm{d}}$ and $n_{e}$. As for the 
shock surface area $A$, each particle represents an area of the shock, since a 
shock front is comprised of particle in a SPH simulation. The shock 
discontinuity is smoothed by the size of the SPH kernel, $h_{\mathrm{SPH}}$, 
which contains $N_{\mathrm{SPH}}$ particles, with a corresponding volume of 
$h^{3}_{\mathrm{SPH}}$. Therefore, one particle belonging to the shock front 
represents a shock area of
\begin{equation}
A\,=\,f_{\mathrm{\small{A}}}\frac{h_{\mathrm{\small{SPH}}}^{2}}{N_{\mathrm{SPH}}}\,,
\label{eq:shock_area_const}
\end{equation}
where $f_{\mathrm{A}}$ is a normalization constant which was set to be $6.5$ by
\cite{hoeft2008} using shock tube simulations. Here, we will correct this 
factor also using shock tubes, as it was found that $f_{\mathrm{A}}$ depends 
strongly on $\mathcal{M}$ for low Mach number shocks (see next section).

\begin {table}
  \begin {center}
    \begin {tabular}{|cccc|}
      \hline
      \hline
      $\mathcal{M}$ & $P_{\mathrm{1}}$ & $\mathcal{M}_{fit}$ & 
      $\sigma(\log{\mathcal{M}})$\\ 
      \hline
      \hline
      1.4  & 992285.25 & 1.38    & 0.003 \\
      1.5  & 789384.98 & 1.48    & 0.003 \\ 
      2    & 348961.73 & 1.97    & 0.003 \\
      3    & 133227.50 & 2.92    & 0.007 \\
      6    & 30602.35  & 5.25    & 0.020 \\
      10   & 10825.70  & 8.61    & 0.023 \\
      30   & 1192.46   & 25.41   & 0.031 \\
      60   & 297.87    & 50.47   & 0.034 \\
      100  & 107.22    & 84.14   & 0.035 \\
      300  & 11.91     & 246.04  & 0.042 \\
      500  & 4.29      & 407.38  & 0.045 \\
      700  & 2.19      & 566.24  & 0.046 \\
      1000 & 1.07      & 803.53  & 0.050 \\
      1300 & 0.63      & 1042.32 & 0.051 \\
      1500 & 0.48      & 1199.50 & 0.051 \\
      1800 & 0.33      & 1419.06 & 0.053 \\
      \hline
      \hline
    \end {tabular}
    \caption{\small{Shock tube tests parameters. The columns indicate the 
strength of the shock, i.e., the Mach number $\mathcal{M}$, the pressure in the
high density region, the fitted Mach number and the standard deviation, 
$\sigma$, of the fitted Gaussian.}}
\label{table:shocktubes}
  \end {center}
\end {table}

\begin{figure*}
\centering
\includegraphics[width=1.0\textwidth]{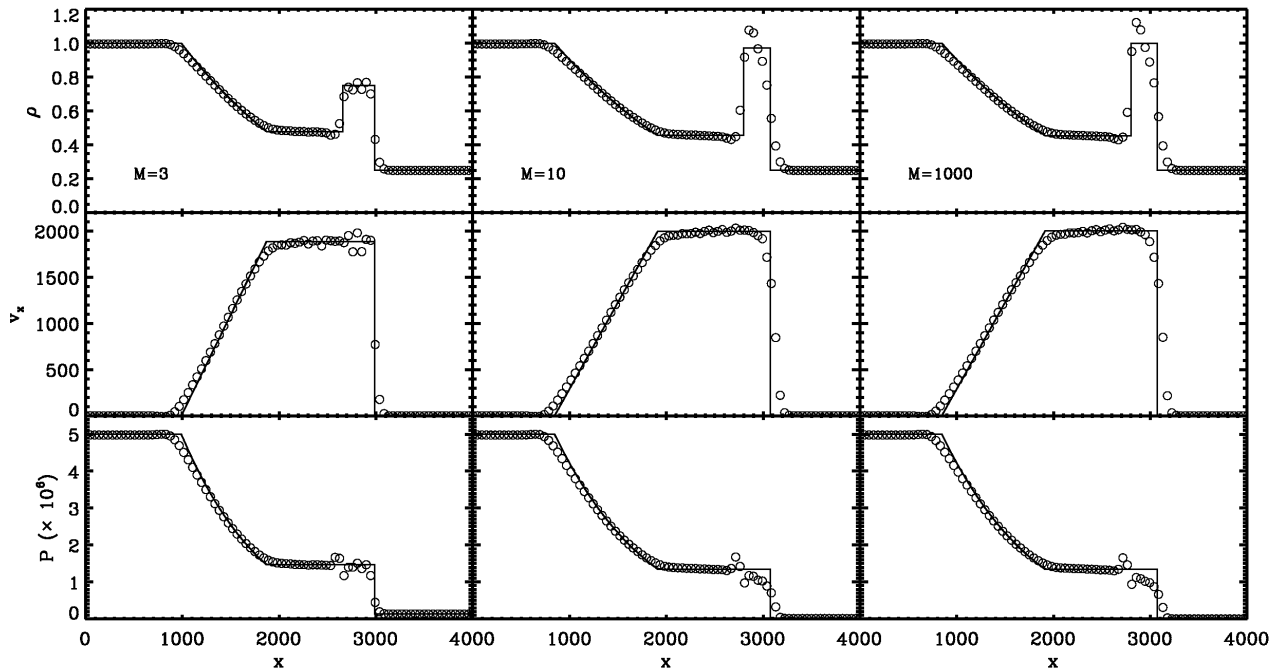}
\caption{\small{Simulations of the Sod shock tube problem with Mach numbers 
$\mathcal{M}=3$, $10$ and $1000$. The numerical result of the volume-averaged 
hydrodynamical quantities within bins with a spacing equal to the inter-particle
separation of the high-density medium is shown in open circles while the 
analytical result is shown as continuous lines. The upper row shows the density
$\rho$ of the gas, the middle row shows the velocity $v_{x}$, while the bottom 
row depicts the pressure of the gas.}}
\label{fig:shocktube}
\end{figure*}
\section{Calibrating the Radio Emission}
\label{sec:shocktubes}

An important quantity in Eqn.~\ref{eq:total_emission2} is the shock surface 
area, which depends on the normalization factor $f_{\mathrm{A}}$. 
\cite{hoeft2008} determined this factor using shock tube tests, finding that it
depends on the Mach number of the shock. The origin of this is the following: 
the Mach number estimator results in a distribution of Mach numbers for the 
particles in the SPH-broadened shock discontinuity, i.e., in the periphery of 
the shock front the Mach number is underestimated. For small Mach numbers, 
$\mathcal{M}\lesssim 5$ the radio emission depends strongly on $\mathcal{M}$,
while for large Mach numbers it does not. Therefore, a constant $f_{\mathrm{A}}$
underestimates the radio emission of shocks with $\mathcal{M}\lesssim 5$.
Next, we determine the function $f_{\mathrm{A}}(\mathcal{M})$ by calibrating it 
with shock tube tests.

We run several shock tube tests in order to model a large range of Mach 
numbers and, therefore, obtaining a proper area function. We construct 16 
standard shock tube tests \citep{sod1978} using a 3D-version of the GADGET2 
code \citep{springel2005}. We consider an ideal gas with $\gamma=5/3$, 
initially at rest. The left-half space ($x<2000$) is filled with gas at unit 
density, $\rho_{2}=1$ and pressure $P_{2}=5\times10^{6}$, while the right half, 
$x>2000$, is filled with low-density gas, $\rho_{1}=0.25$ and variable 
low-pressure. The value of this low-pressure gas has been chosen such that the 
resulting solutions yield the Mach numbers in the range of 
$\mathcal{M}=1.4-1800$.
%[1.4, 1.5, 2, 3, 6, 10, 30, 60, 100, 300, 500, 700, 1000, 1300, 1500, 1800]$. 
We set up the initial conditions in three dimensions using an 
irregular glass-like distribution. In order to test the differences in 
resolution, we ran the same tests in different resolutions: from $12 500$
particles to $800 000$ particles. We did not find any significant difference 
between these resolutions. Hence, the results presented here are done with
shock tubes with $100 000$ particles. 
The particles are contained in a periodic box which is longer in the 
$x$-direction than in the other two dimensions, $y$ and $z$. The parameters of
the different shock tube tests are shown in Table~\ref{table:shocktubes}.

Fig.~\ref{fig:shocktube} shows the profile of gas density, velocity and pressure
in a shock tube calculation where the gas particles experience a shock of Mach
number $\mathcal{M}=10$. The simulation results are presented by circles and
the continuous lines give the analytic solution. The shock tube simulations were
run with the same SPH parameters (e.g., artificial viscosity) as the 
MareNostrum Universe in order to compare results. Overall, there is a good 
agreement between the analytical and the numerical solution, with 
discontinuities resolved in about two to three SPH smoothing lengths. There is 
a characteristic pressure blip at the contact discontinuity. The irregularities 
observed in the density profile may be due to the small number of particles in 
the SPH kernel, namely $N_{\mathrm{SPH}}=40$.  

\begin{figure}
\centering
\includegraphics[width=0.40\textwidth]{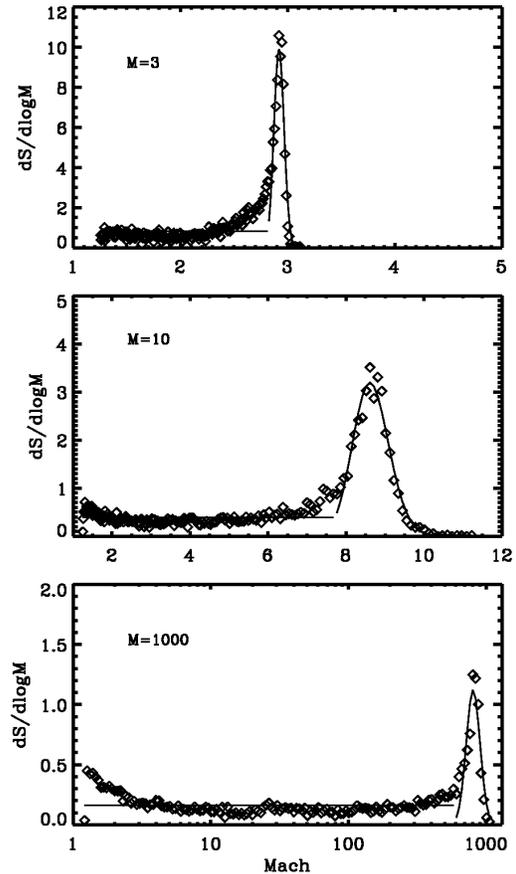}
\caption{\small{Differential shock surface area as a function of logarithmic
Mach number bins for a shock tube calculation with $\mathcal{M}=10$. The
differential shock surface area is normalized by the binsize and the cross 
section of the tube. Solid line represents the bestr fit of the Gaussian-like
distribution and the constant surface area to the left of the Gaussian.}}
\label{fig:areatube}
\end{figure}

We calculate the strength of the shock with the Mach number estimator. Although
the shock tube was run with $N_{\mathrm{SPH}}=40$, the shock finder was set up 
with $N_{\mathrm{SPH}}=80$ in order to smooth the fields. To determine the
function $f_{\mathrm{A}}(\mathcal{M})$, we calculate the differential shock 
surface area in logarithmic Mach number bins, $dS/d\log{\mathcal{M}}$, for each 
of the shock tubes. Fig.~\ref{fig:areatube} shows such distribution for the 
shock tube calculation with $\mathcal{M}=10$. The surface area was calculate 
using Eqn.~\ref{eq:shock_area_const} setting $f_{\mathrm{a}}=1$ in order to 
obtain $f_{\mathrm{a}}(\mathcal{M})$. Visual inspection allow us to draw two 
observations. Firstly, there is a small shift in the value of the theoretical 
Mach number and the one estimated by the shock finder meaning the shock finder 
is underestimating the Mach number. Secondly, there is a ``tail'' of shock 
surface areas to the left of the Gaussian-like distribution. This has the 
following origin: the shock finder identifies low-Mach number shocks in the 
periphery of the transition region between the upstream and the downstream zone
(see Fig.~\ref{fig:shocktube}) given the nature of SPH, i.e., quantities are 
calculated over a region determined by $N_{\mathrm{SPH}}$. Hence, the tail has 
surface area contribution from the upstream and downstream region.

We calculate the shock surface area of all shock tube realizations. Every shock
tube exhibits a similar $dS/d\log{\mathcal{M}}$ distribution as the one shown in
Fig.~\ref{fig:areatube}. We fit the distribution with a Gaussian centered
on the mean Mach number $\bar\mathcal{M}$ (which should be the theoretical
Mach number) and a Heaviside function $\mathcal{H}$ multiplied by a constant 
factor $\mathcal{K}$ which is determined as the mean value of the surfaces 
areas located approximately $3\sigma$ to the left of the Gaussian mean value, 
i.e.,

\begin{equation}
f(\log{\mathcal{M}})\,=\,\left\{
\begin{array}{rrl}
\mathcal{K}\cdot\mathcal{H}(\mathcal{X})&\;\;\mathrm{for}&%\;\;
\log{\mathcal{M}}\leq\log{\bar{\mathcal{M}}}-3\sigma\\%\nonumber
e^{-\mathcal{X}^{2}/2\sigma^{2}}&\;\;\mathrm{for}&
\log{\mathcal{M}}>\log{\bar{\mathcal{M}}}-3\sigma
\end{array} \right.
\label{eq:functionmach}
\end{equation}
where $\mathcal{X}=\log{\mathcal{M}}-\log{\bar{\mathcal{M}}}$.

\begin{figure}
\centering
\includegraphics[width=0.32\textwidth]{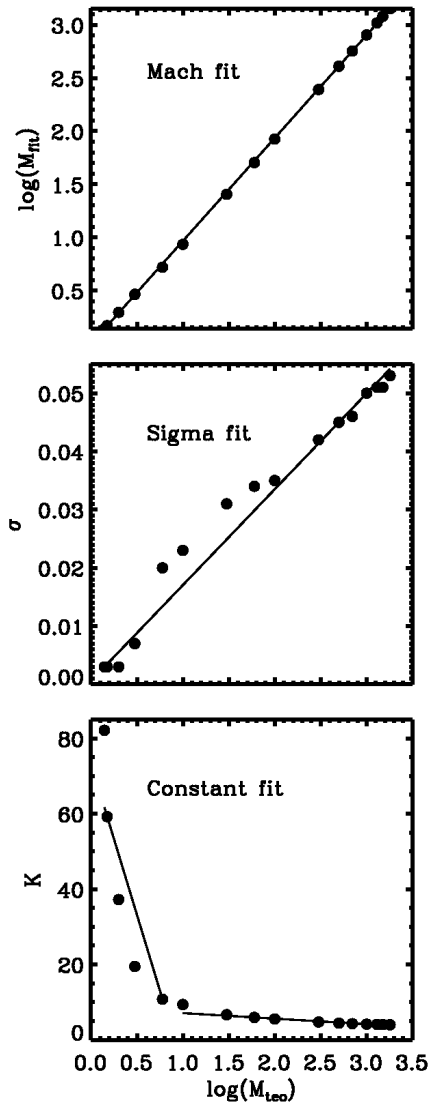}
\caption{\small{Best fits to the different variables of the Gaussians. Upper 
panel: linear fit of the measured Mach number. Central panel: linear fit of the
$\sigma$ of the Gaussian distribution. Bottom panel: linear fit to the constant
$\mathcal{K}$ left to the Gaussians fits.}}
\label{fig:fitpar}
\end{figure}

Each shock tube gives us 3 different fitting values: the constant $\mathcal{K}$,
the mean of the Gaussian function $\bar{\mathcal{M}}$ and the standard deviation
of the Gaussian, $\sigma$. Fig.~\ref{fig:fitpar} shows the best fits to the
mentioned variables. Although we were able to fit the constant $\mathcal{K}$ 
in two segments of Mach numbers, we find it is, on average, $\sim 10\%$ of the 
peak value of the Gaussians. For $\mathcal{M}$ and $\sigma$, we see there is a
good correspondence with the theoretical Mach number. We can, therefore, do
a linear fit, obtaining functions that will deliver a correct Mach number
$\mathcal{M}$ that will be applied to the radio emission model.

\subsection{Calculating the Shock Area}
\label{sec:shockarea}

To correctly calculate the radio emission, we need to correct the normalization
constant $f_{\mathrm{A}}(\mathcal{M})$ (see Eqn.~\ref{eq:shock_area_const}).To 
this end, we find the area within 2$\sigma$ around the mean Mach number of each
shock tube, $S_{\mathrm{exp}}$. This area should be similar to the cross section 
of the tube, $S_{\mathrm{theo}}$. We find a relation between the Mach number and 
the ratio $S_{\mathrm{theo}}/S_{\mathrm{exp}}$:
\begin{equation}
f_{\mathrm{A}}(\mathcal{M})\,=\,0.54\log{\mathcal{M}}\,+\,5.71\,.
\label{eq:fa}
\end{equation}

The fitting obtained in the previous section together with Eqn.~\ref{eq:fa}, 
allows us to correctly estimate the radio emission 
(Eqn.~\ref{eq:total_emission2}) in the simulation.

Another important quantity we want to determine from the MareNostrum 
simulation is the shock surface area in a given logarithmic Mach number 
interval. However, as seen in the previous section, our shock finder algorithm 
underestimates high Mach numbers and, due to the SPH nature, for high Mach 
numbers it identifies low shock Mach numbers to particles involved in the shock
front. In order to correct this, we will use the results of the previous section
and deconvolve them with the measured $dS/d\log{\mathcal{M}}$ distribution, 
i.e., we will deconvolve the shock surface area distribution calculated from 
the MareNostrum simulation with a deconvolution kernel obtained by constructing
the function of Eqn.~\ref{eq:functionmach}. We do this by using a deconvolution
technique which is explained in Appendix~\ref{app:dsdm}. 

\begin{figure}
\centering
\includegraphics[width=0.48\textwidth]{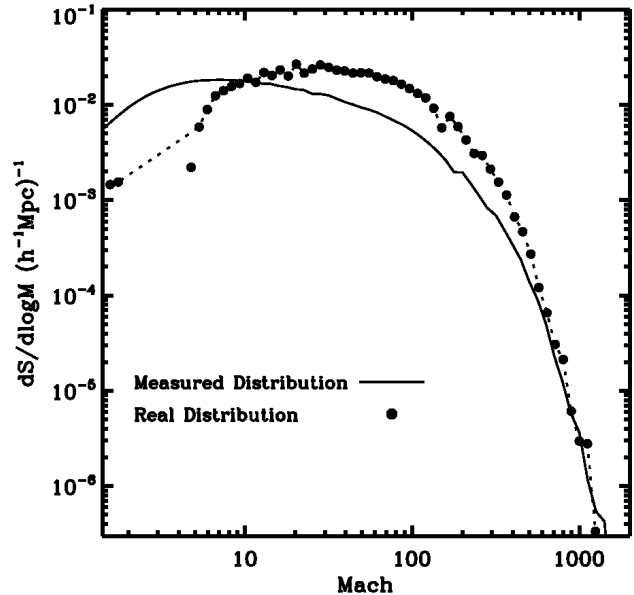}
\caption{\small{Shock surface area in logarithmic bins as a function of Mach 
number. Solid line depicts the measured distribution obtained from the 
MareNostrum simulation. Filled circles shows the ``real'' distribution, i.e.,
the distribution obtained by deconvolving the measured distribution with the
data from the shock tubes.}}
\label{fig:corrected_dist}
\end{figure}

Fig.~\ref{fig:corrected_dist} shows the measured $dS/\log{\mathcal{M}}$ 
distribution obtained from the MareNostrum simulation (solid line) and the 
``real'' distribution obtained with the deconvolution technique (filled circles 
linked by a dotted line), where we have also used the Mach number shift 
obtained from fitting $\bar{\mathcal{M}}$ from the shock tubes. As expected 
from the shock tube data, the real distribution has lower Mach numbers, 
expressed in a smaller shock surface area, while it has more high Mach numbers 
for $\mathcal{M}\gtrsim 10$. 

\begin{figure*}
\centering
\includegraphics[width=0.90\textwidth]{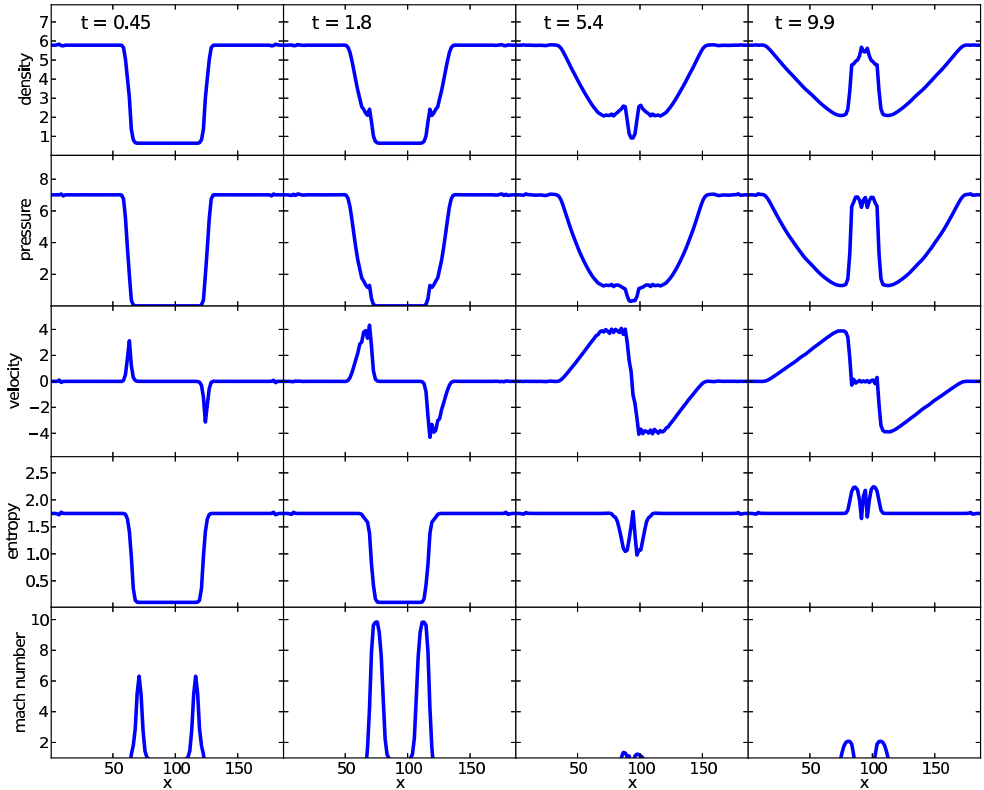}
\caption{\small{Simulation of two cooling shocks. The initial densities and 
pressures are chosen in way that two shock fronts each with Mach number 10 
forms when the simulation starts.}}
\label{fig:colliding_shocks}
\end{figure*}

The Mach number determination is constructed in a way that it results in a
conservative estimate even in more complex situations. We would like to 
demonstrate this by a colliding shock tube simulation. We have set up a shock 
tube with three initial zones, with densities and internal energies so that two
shocks with Mach number 10 forms later on, see Fig.~\ref{fig:colliding_shocks}.
Our shock finder detects the shock fronts even when the actual shock is not 
spatially separated from the initial pressure jump. In the shock finder the 
Mach number is estimated via two methods: firstly it evaluates the velocity 
field and secondly it uses the entropy jump.  In the initial set-up the entropy
jump would lead to a Mach number of 10, corresponding to the actual Mach 
number. However, particles needs to be accelerated to form the corresponding 
velocity field. Therefore the shock finder underestimates the Mach number in 
the beginning of the simulation. Later on, when the two shock fronts get close 
to each other, the shock finder includes particles from the downstream region 
of the oncoming shock, hence,  it does not measure the actual upstream 
properties anymore. A mach number estimate based on the velocity field would 
significantly overestimate the Mach number, since the detected velocity 
divergence would be spuriously high. However, the entropy now causes that the 
shock finder underestimates the Mach number. Therefore, our shock finder tends 
to provide only a lower estimate for the Mach number in complex situations, 
where e.g. several shock fronts overlap. 
 
\subsection{Correcting the floor temperature}
\label{sec:floortemperature}

\begin{figure}
\centering
\includegraphics[width=0.50\textwidth]{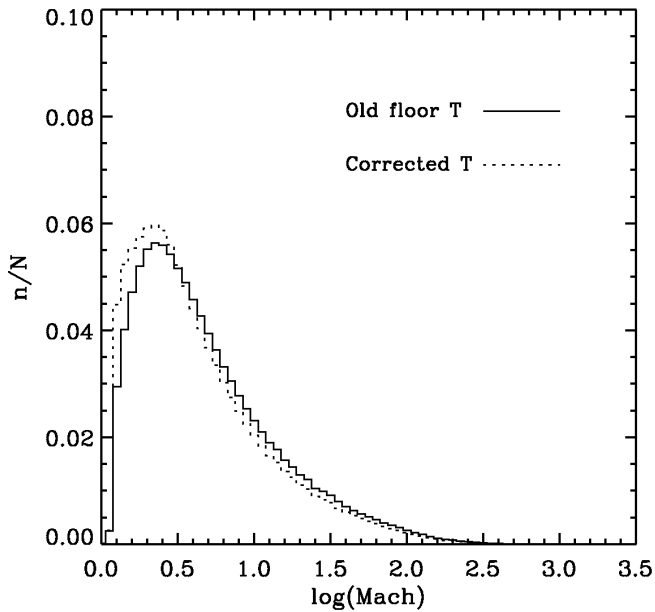}
\caption{\small{Histogram showing the Mach number distribution when the original
MareNostrum floor temperature is used (solid line), and when applying Eqn.~ 
(dotted lines).}} 
\label{fig:machhistogram}
\end{figure}

One last step is to correct for the floor temperature of the MareNostrum 
simulation. Mach numbers are sensitive to the temperatures of the different 
environments. Since the simulation was run with a low floor temperature (200 K)
and without any UV background,  expansion cooling leads to very low 
temperatures. This can result in unrealistically high Mach numbers. However, 
the UV background heating leads to a simple power law between the temperature 
and the density, so the temperature of cold regions (underdense regions) can be
scaled by 
\citep{hui1997}
\begin{equation}
T\;=\;T_{0}(1+\delta)^{\gamma-1}\,,
\label{eq:floortemp}
\end{equation}
for regions with $\delta\lesssim 5$, where $\delta$ is the density contrast,
and temperature $T_{0}$. We apply this relation to low-density particles. 
Fig.~\ref{fig:machhistogram} shows the Mach number distribution with the
``old'' floor temperature (solid lines) and with the corrected temperatures. 
Two effects can be seen: i) by correcting the temperature, more particles get 
assigned a Mach number, and ii) the values of the Mach numbers decreases, i.e.,
UV background heating plays an important role in cosmological shocks. 
Nonetheless, if we compare our results to those of \cite{skillman2008}, there
still high Mach numbers. In their simulations, they did not find shocks 
above $\mathcal{M}\approx 200$. This may be due to the fact that the floor 
temperature was too low, so any scaling to those low temperatures, which, in 
general, are in underdense regions, will result in a low increment of the 
temperature. In the case of \cite{skillman2008}, their floor temperature was set
at $T=10^{4}$ K, thus assuming the low-density gas to be ionized.

The results in the following sections are done using the corrected temperature
values.

\section{Results}
\label{sec:results}

\subsection{Shock Frequency in the Cosmic Web}
\label{sec:shockfreq}

We computed the surface area of identified shocks per logarithmic Mach number
interval, $dS(\mathcal{M})/\log{\mathcal{M}}$ as described in 
Section~\ref{sec:shockarea}. We divide the shock surface area by the volume 
of the simulation box, as done in \cite{ryu2003}. The inverse of the shock 
surface area, $1/S$, can be of thought as a mean separation of shocks because 
it is the simulation volume divided by the total shock surface area. The 
$dS(\mathcal{M})/\log{\mathcal{M}}$ distribution also indicates the frequency 
with which shocks happen in the Universe.

Instead of studying and classifying shocks as internal or external as done by
\cite{ryu2003}, we use the SpineWeb technique (see 
Section~\ref{sec:spineweb}) in order to characterize voids, walls, filaments
and clusters by their morphology. \cite{aragoncalvo2010b} found that while all
morphologies occupy a roughly well-defined range in density, this is not 
sufficient to differentiate between them given their overlap. Environment
defined only in terms of density fails to incorporate the intrinsic dynamics of
each morphology. Nevertheless, we also construct the differential shock surface
area distribution by using density and temperature cuts in order to compare 
with \cite{skillman2008}.

\begin{figure*}
\centering
\includegraphics[width=0.38\textwidth]{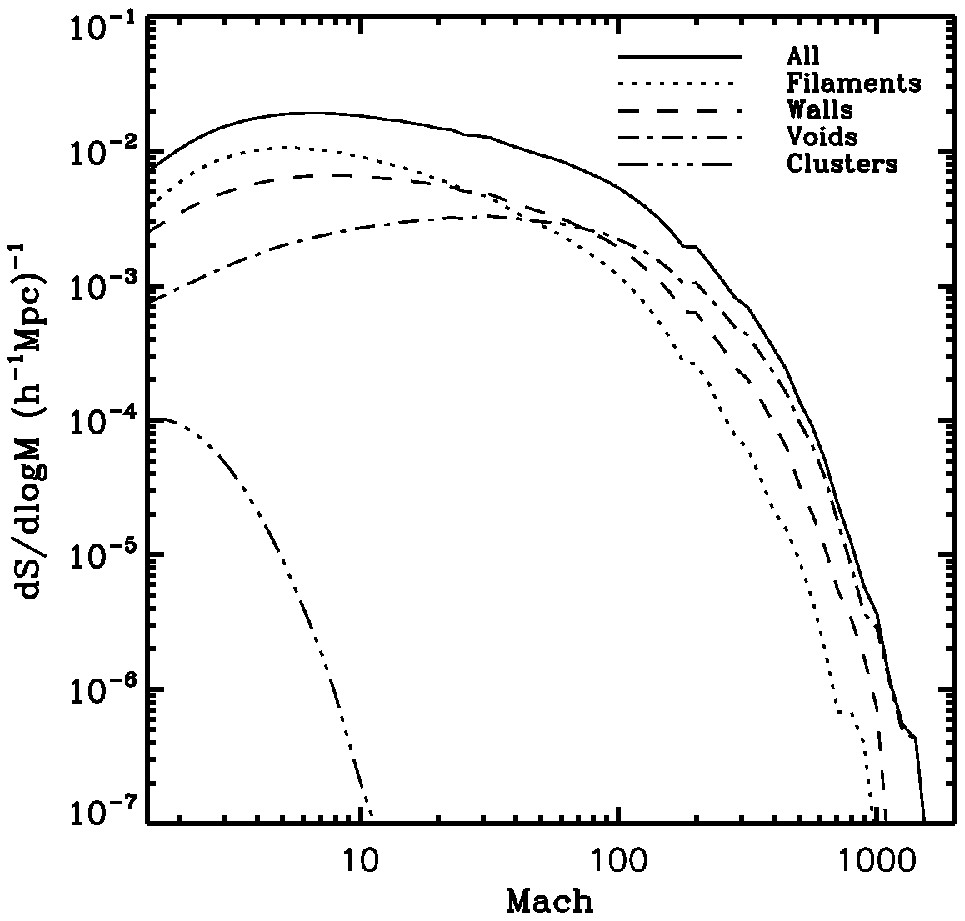}
\includegraphics[width=0.38\textwidth]{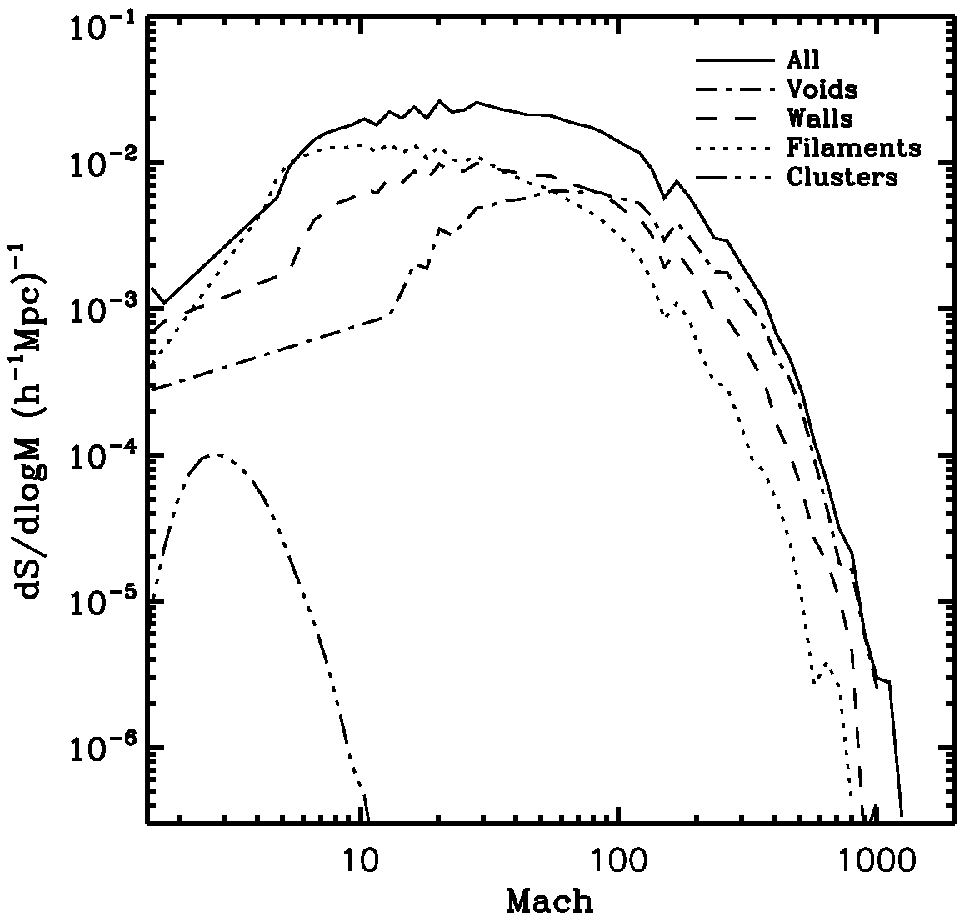}
\includegraphics[width=0.38\textwidth]{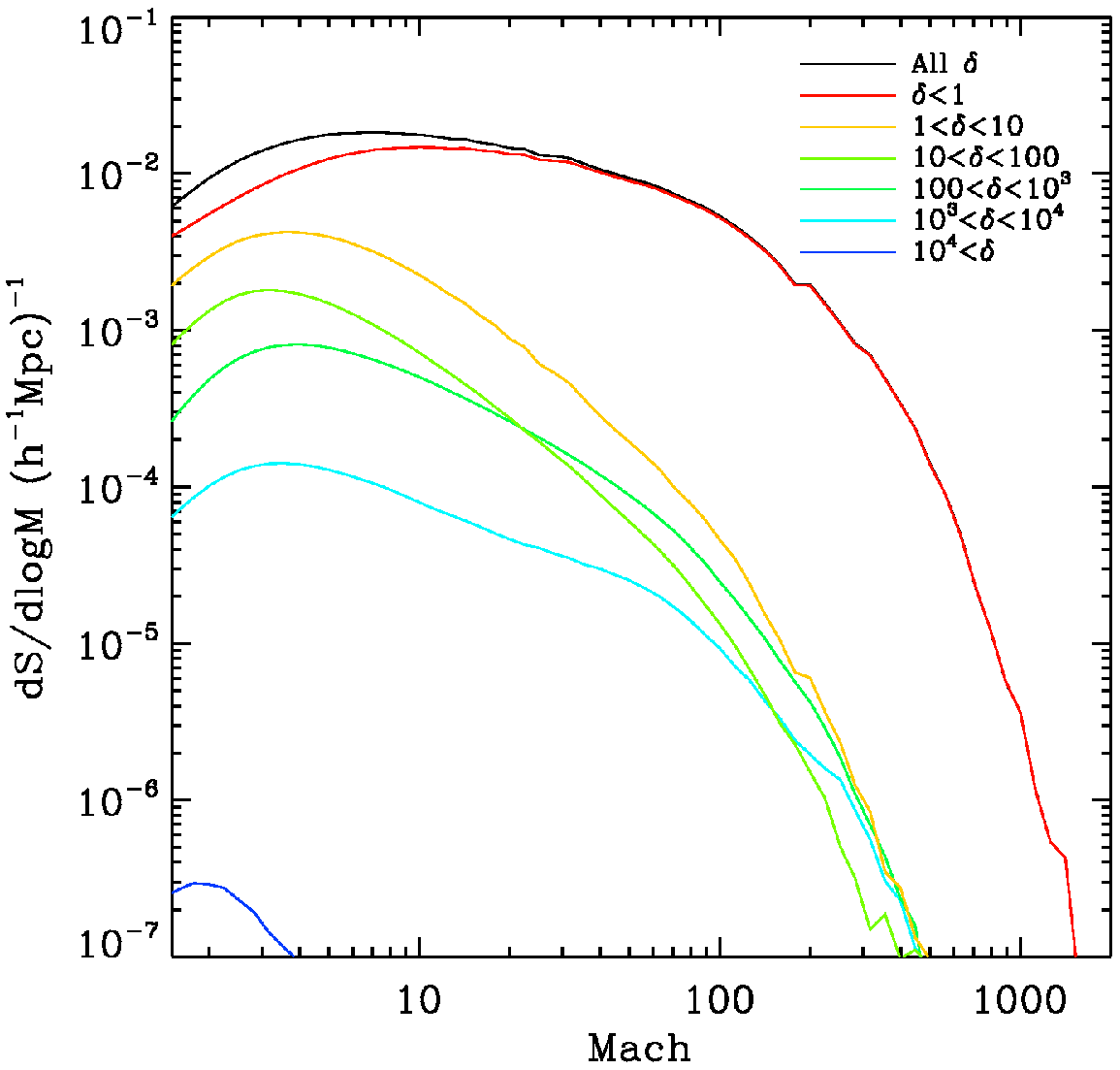}
\includegraphics[width=0.38\textwidth]{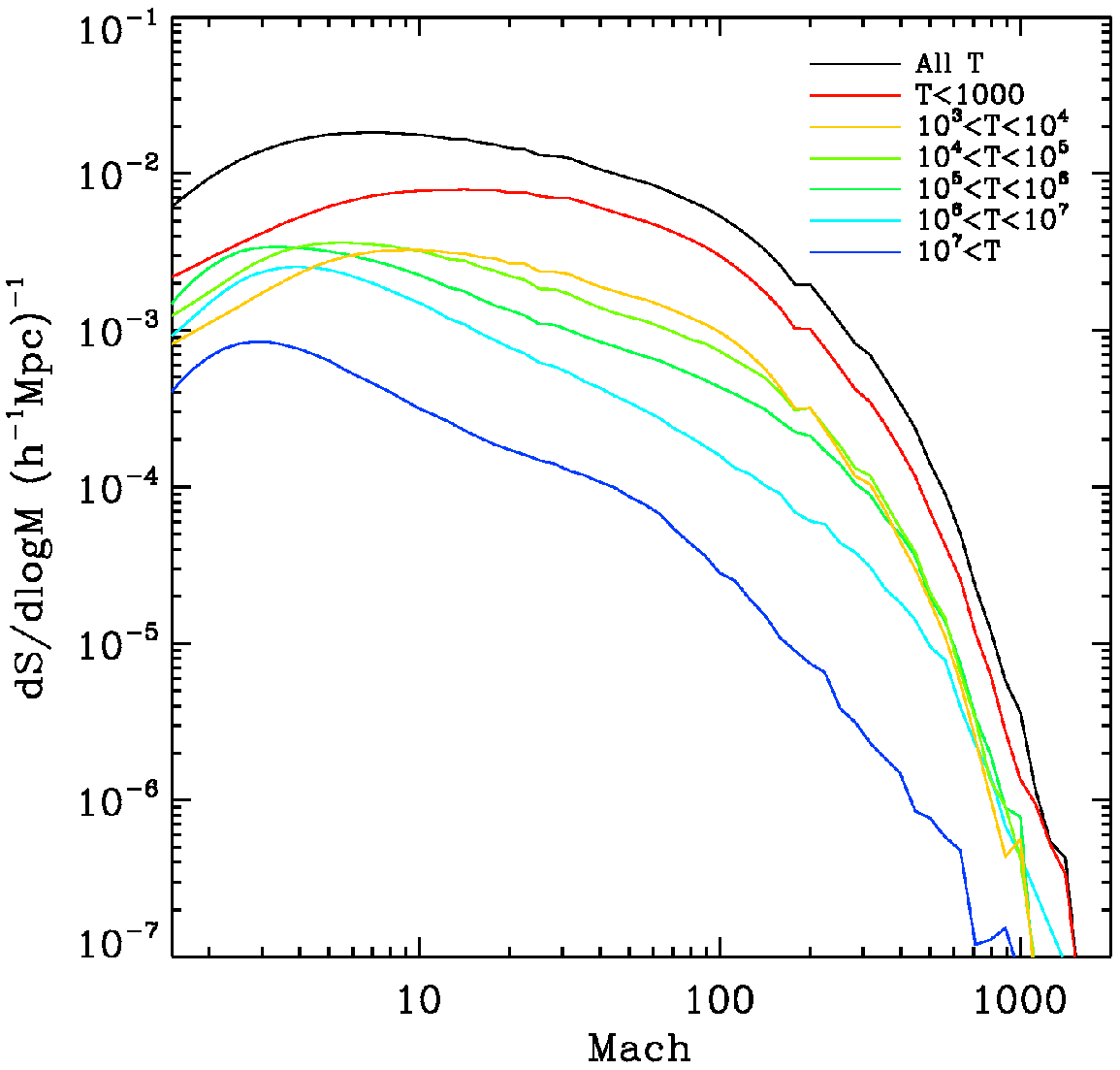}
\caption{\small{Differential shock surface area as a function of logarithmic 
Mach number bins. The top left panel shows the distribution for the different
elements of the cosmic web: voids (dash-dotted line), walls (dashed line), 
filaments (dotted line) and clusters (dash-dot-dotted line). The solid line
shows the combine distribution of all the elements of the cosmic web. By 
applying the deconvolution technique (Sect.~\ref{sec:shocktubes}) to these 
distributions, we obtained the results presented in the top right panel. The 
bottom left panel shows the shock surface distribution divided into several 
ranges of overdensity. The bottom right panel shows the distribution divided 
into several temperature ranges.}}
\label{fig:shocksurf}
\end{figure*}

Fig.~\ref{fig:shocksurf} shows the differential shock surface area as a 
function of logarithmic Mach number bins. The top left panel shows the shock 
surface area of the different elements of the cosmic web as computed by
the SpineWeb procedure. The top right panel shows the latter results after
deconvolving them with the method outlined in Appendix~\ref{app:dsdm}. The 
bottom left panels shows the distribution divided into several ranges of 
overdensity values, while the bottom right panel shows the distribution divided
in temperature ranges.

We see that there is a large range of shocks of different strengths, with Mach
numbers in the range $\mathcal{M}=1.5\sim 1700$. However, it is important to
notice that the large Mach numbers may not occur in reality, and even the 
correction applied to the temperatures in the low-density regions (see
Section~\ref{sec:floortemperature} may not be enough to correctly correct the
temperatures in the low density regions.

When looking at the top panels, we observe that each element of the cosmic web 
has a distinctive shock surface area distribution, which follows their 
characteristic density and temperature distribution and the dynamics between 
them. After deconvolving the shock surface area distribution obtained from the
MareNostrum simulation using the method outlined in Appendix~\ref{app:dsdm}, we
see that there is a shift in the distribution. In order to quantify this, we 
will define the characteristic Mach as the median Mach number of the different
distributions. For example, for clusters, after deconvolving the shock surface 
area distribution with the deconvolution kernel, the characteristic Mach number
is $\mathcal{M}=1.8$, slightly different from the value of $\mathcal{M}=1.6$ 
obtained without deconvolution. The next analysis is done using the 
deconvolution results.

Shocks in voids have a low surface area for low Mach numbers in the range 
$\mathcal{M}\approx 1.5-10$. The distribution shows a peak at 
$\mathcal{M}\approx 100$, extending up to $\mathcal{M}\approx 1300$. 
The characteristic Mach number is $\mathcal{M}_{\mathrm{voids}}\approx 18$. It is 
expected that voids have strong shocks since void regions have low temperatures
and they accrete onto walls and/or filaments, which are denser, hotter 
structures. Walls present a similar distribution than that of voids, but they 
have a higher shock surface area for $\mathcal{M}\lesssim 10$, and then 
decreases for larger Mach numbers. The characteristic Mach number of walls is 
$\mathcal{M}_{\mathrm{walls}}\approx 7.5$.

On the other hand, filaments present a characteristic Mach number
$\mathcal{M}_{\mathrm{filaments}}\approx 6.2$, in which their surface area is 
higher than that of voids and walls. The surface area then decreases at 
$\mathcal{M}\approx 40$, becoming less than that of voids and walls.

Clusters have, in comparison with the rest of the cosmic web, weak shocks, with
a maximum Mach number of $\mathcal{M}\approx 10$. The shocks that corresponds 
to the cluster environments are interior cluster shocks. In this region, 
temperatures are high and homogeneous, without large temperature jumps. The 
characteristic Mach number for interior cluster shocks is 
$\mathcal{M}\approx 1.8$.

We also construct the differential shock surface area in terms of temperature
and density, in order to compare with \cite{skillman2008}. This is shown in
the lower panels of Fig.~\ref{fig:shocksurf}. The difference between these
panels and the upper panels is that here we divide the shock surface area in
temperature and density ranges. The temperature and density ranges are intended
to denote the different temperature and density of the elements of the cosmic
web. for simplicity, we do not deconvolve these distributions.
When looking at density cuts (lower left panel of Fig.~\ref{fig:shocksurf}), it 
is interesting to notice that cluster cores ($\delta>10^{4}$) have low Mach 
numbers and low surface area. This is mainly due to the fact that this region 
has already been heated, and so its temperature is high. Low-density regions 
($\delta <1$) may be associated with voids. These regions have high 
characteristic Mach numbers due to their low density and low temperature. We 
see that there is an increase of the characteristic Mach number as the density 
decreases.

Similar results are observed when looking at the temperature cuts. Each cut 
has a characteristic Mach number, which increases as temperature decreases. 
For high temperature ($T>10^{7}$ K), the characteristic Mach number is 
$\mathcal{M}_{\mathrm{char}}\approx 2$, while for $T<1000$ K, 
$\mathcal{M}_{\mathrm{char}}\approx 20$. \cite{skillman2008} states that this
characteristic Mach number is due to the maximum temperature jump possible for
a given temperature.

%In general, we find good agreement with the result of \cite{skillman2008} for 
%cuts with low density and temperature. For dense structures, e.g., clusters, we
%see that our distribution is similar to low density regions, even extending to
%high Mach numbers. This is not observed in the results of \cite{skillman2008}.
%This may be an effect of SPH, since quantities are obtained by averaging over
%a kernel, thus a particle that is in a low density region is assigned a high
%Mach number (see Sect.~\ref{sec:shocktubes}). This does not happens when 
%identifying structures in terms of their morphology. Hence, although 
%it is possible to get a rough estimation of shock distribution in the cosmic
%web by a simple classification scheme based on overdensity, a 
%morphologically-based measure quantifies the precise location of the 
%cosmological formation of shocks.

In general, we find good agreement with the result of \cite{skillman2008} for 
cuts with low density and temperature. For dense structures we find that our 
results differ significantly from \cite{skillman2008}. More precisely, we find 
that even in regions with $\delta > 10^3$ shocks with $\mathcal{M} \gtrsim 100$
can be found, while  \cite{skillman2008}  find that shocks have Mach numbers 
basically below 10 for regions with $\delta > 10^3$. Comparing the density cuts
(lower left panel of Fig.~\ref{fig:shocksurf}) with the upper panels we see 
that dense regions with high Mach number does not reside in clusters. Instead, 
these are dense structures in filaments, walls or voids. The particle based 
TreeSPH simulation technique, as used in Gadget, tends to from clumpy 
structures. Therefore, we find dense structures with $\delta > 10^3$ even 
outside of clusters. If these structures reside close to an accretion shock, we
find a high Mach number at a relatively high density. This effect is enhanced by
smearing out shock fronts in SPH by approximately two times the smoothing 
length.

For region with $\delta > 10^4$ we find significantly less shocks than 
\cite{skillman2008}. The difference is likely caused by a lower effective 
resolution in our SPH simulation. \cite{skillman2008} used a simulation box 
with a side length of $512 \, h^{-1} \, {\rm Mpc}$. They used a root grid 
resolution for both dark mater and gas of 512. For dark matter the MareNostrum 
simulation is significantly better resolved. In the AMR scheme with a density 
refinement criterium the resolution follows the mass roughly in a similar way 
as SPH. However, since the SPH kernel smoothes over $N$ particles (64 in our 
case), the effective resolution for gas is in the MareNostrum simulation by a 
factor of 2 lower than in the 'Santa Fe Light Cone' used by 
\cite{skillman2008}. A higher resolution as used in the MareNostrum simulation 
is necessary to form shock fronts in regions with $\delta > 10^4$.

\subsection{Temperature Distribution and Physical Properties of the Cosmic Web}

As mentioned earlier, one of the big advantages of the SpineWeb procedure is 
that we can precisely characterize the different elements of the cosmic web. 
We can, therefore, study the temperature and density distribution in the 
different elements of the cosmic web.

\begin{table}
 \caption{Mean temperature and mean density contrast of the different elements 
of the cosmic web at $z=0$.}
 \begin{minipage}{0.95\linewidth}
   \centering
   \begin {tabular}{|l||c|c|c|c|}
     \hline
     & & \\
     & $\langle\mathrm{T}\rangle$ (K) & $\sigma_{T}$ & $\langle\delta\rangle$ &
     $\sigma_{\delta}$\\
     & & \\
     \hline
     \hline
     & & \\
     Clusters & 7.90$\times10^{7}$ & 8.13$\times10^{7}$ & 885.73 & 1108.28 \\
     Filaments& 7.30$\times10^{6}$ & 1.44$\times10^{7}$ & 71.87  & 180.27  \\
     Walls    & 2.21$\times10^{6}$ & 6.84$\times10^{6}$ & 18.71  & 71.24   \\
     Voids    & 6.96$\times10^{5}$ & 4.30$\times10^{6}$ & 2.89   & 25.17   \\
     & & \\
     \hline
   \end {tabular}
 \end{minipage}
 \label{tab:table_temp_dens}
\end {table}

Table~\ref{tab:table_temp_dens} shows the mean values of the temperature and
density contrast of the MareNostrum simulation. The shown temperature is the 
simulation temperature corrected using the relation in Eqn.~\ref{eq:floortemp}.
The mean values were obtained by averaging the temperature and density of the
particles that belong to the different environments. As expected, clusters are 
the hottest structures in the Universe, with a mean temperature of 
$7.90\times10^{7}$ K. This is because clusters have deep potential wells into 
which baryons accrete, thus heating the clusters.

The most salient features of the cosmic web are the large filamentary networks
which are interconnected across vast distances. The filamentary network 
permeates all regions of space, even the underdense voids. We find that 
filaments have a mean temperature of $7.30\times10^{6}$ K, which is well within
the range estimated for the filamentary WHIM temperature ($10^{5}-10^{7}$ K).

\begin{figure}
\centering
\includegraphics[width=0.50\textwidth]{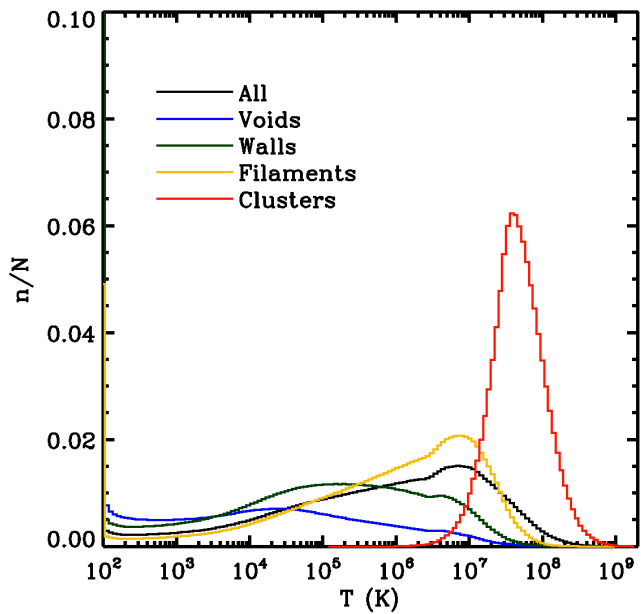}
\caption{\small{Temperature distribution of the different environments of the 
Cosmic Web.}}
\label{fig:temp_distr}
\end{figure}

Table~\ref{tab:table_temp_dens} also shows the density contrast of the 
constituents of the cosmic web. Filaments are the second densest objects in the
Universe, with $\langle\delta\rangle\sim70$. Walls are less dense than 
filaments, with $\langle\delta\rangle~\sim19$. Finally, voids have a mean 
density contrast of $\langle\delta\rangle~\sim3$. 

\begin{figure}
\centering
\includegraphics[width=0.50\textwidth]{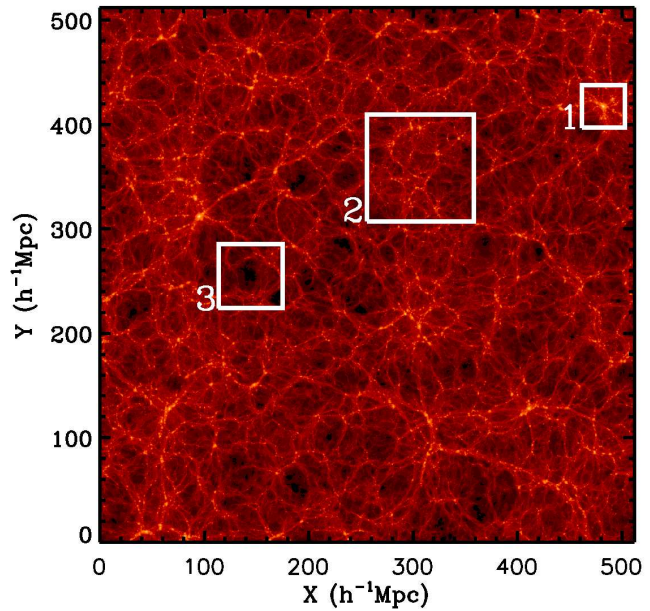}
\caption{\small{Density field of the MareNostrum Universe. The figures 
represents a slice of 20$h^{-1}$Mpc thick through the z-axis. Three elements
are distinguished: the most massive cluster in the simulation (box number 1),
a region of rich filaments (box number 2) and a void region (box number 3). 
The different environments were find using the SpineWeb technique.}}
\label{fig:cosmicweb}
\end{figure}

\begin{figure}
\centering
\includegraphics[width=0.30\textwidth,angle=270]{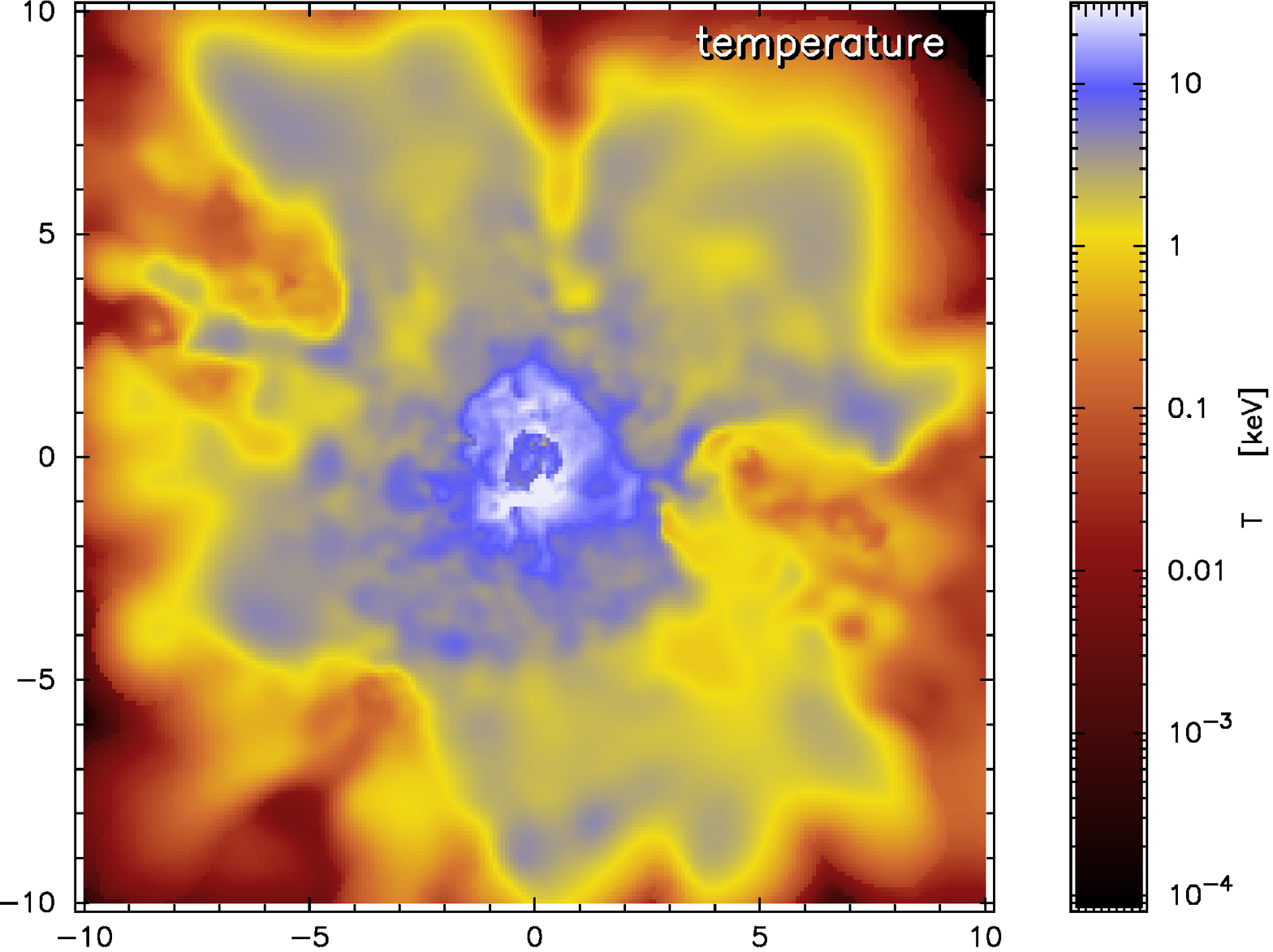}
\includegraphics[width=0.30\textwidth,angle=270]{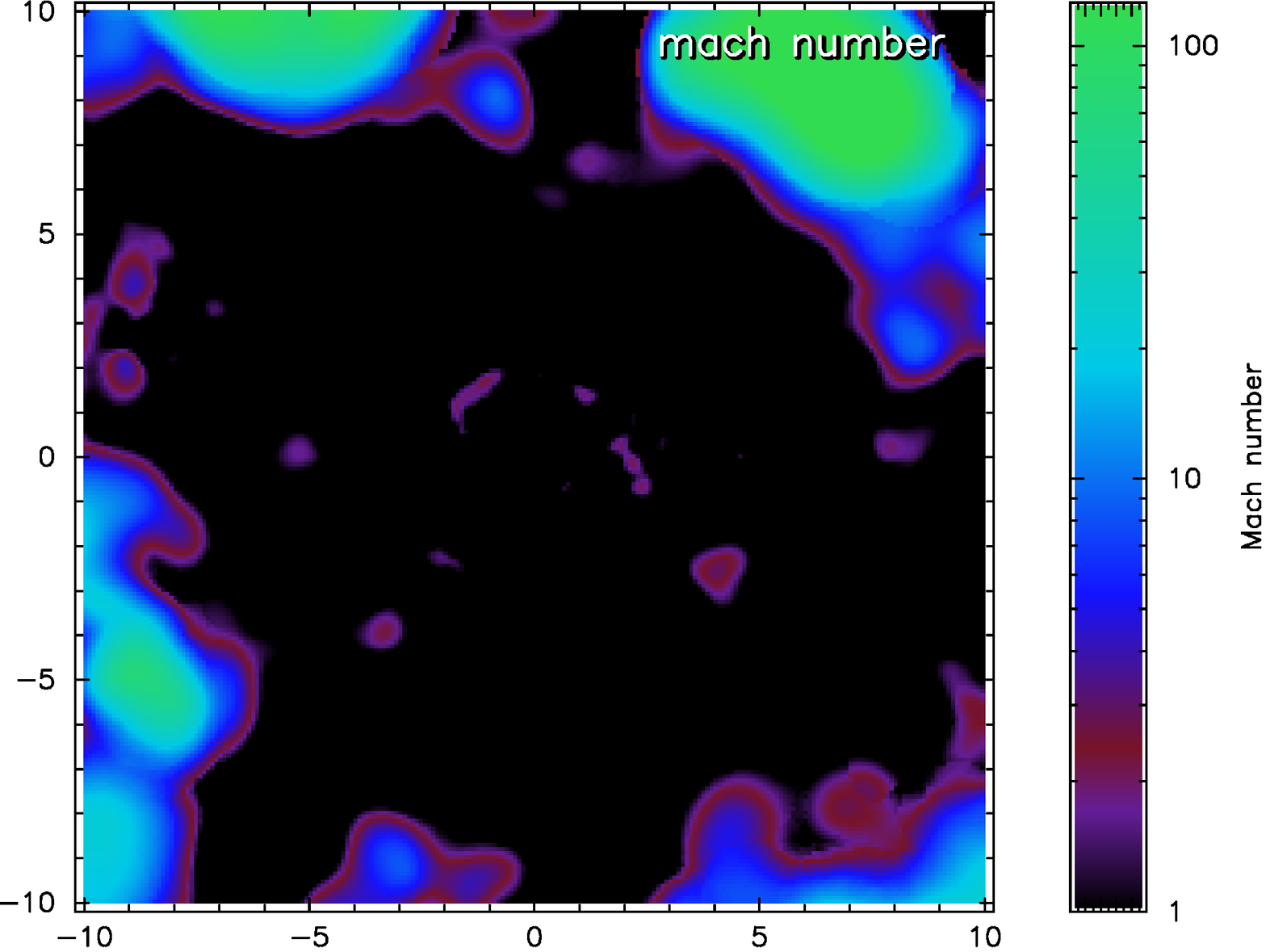}
\caption{\small{Temperature and Mach number in an infinitesimally slice for the
most massive cluster in the MareNostrum simulation (box 1 in 
Fig.~\ref{fig:cosmicweb}).}}
\label{fig:massivecluster}
\end{figure}

\begin{figure}
\centering
\includegraphics[width=0.43\textwidth]{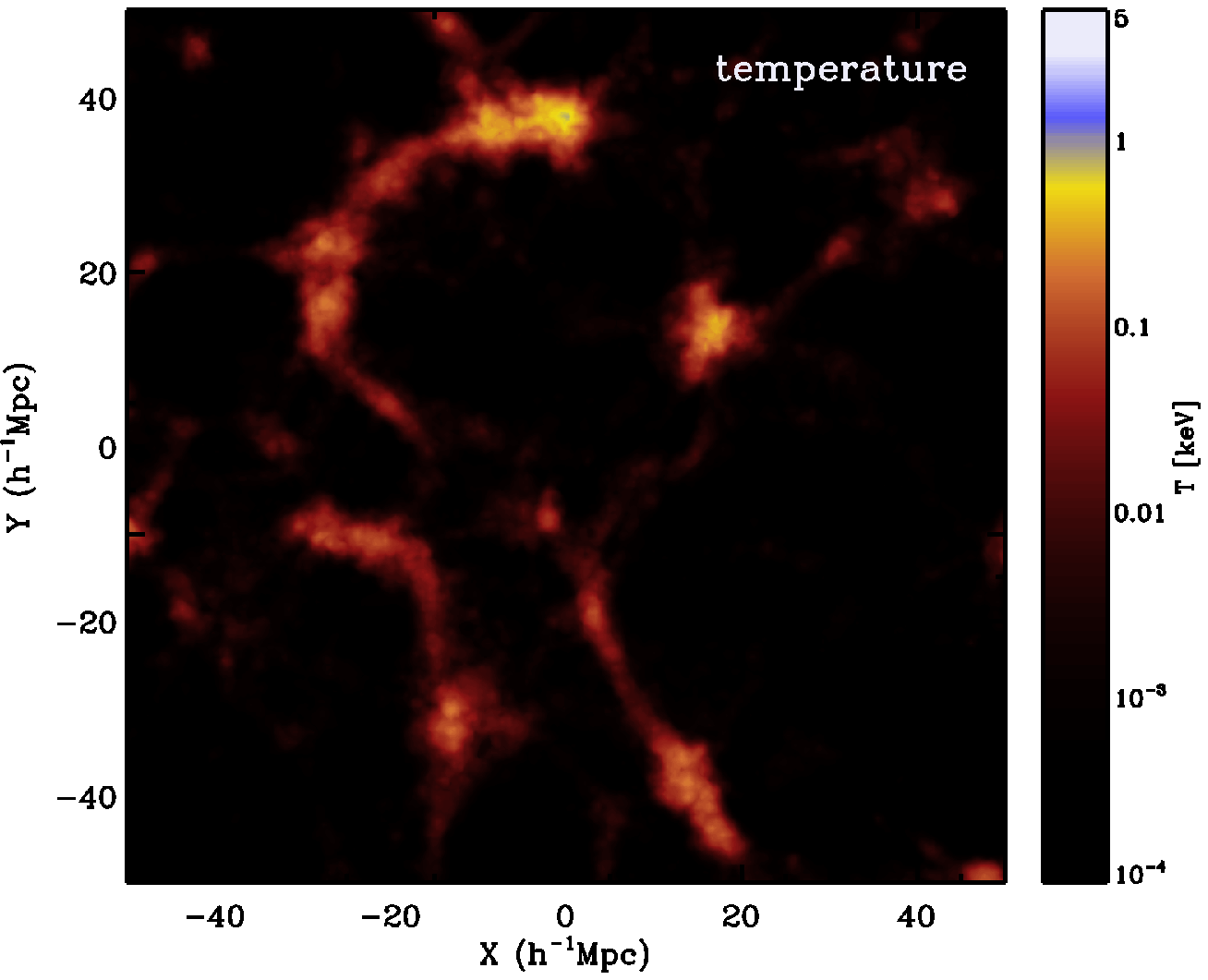}
\includegraphics[width=0.43\textwidth]{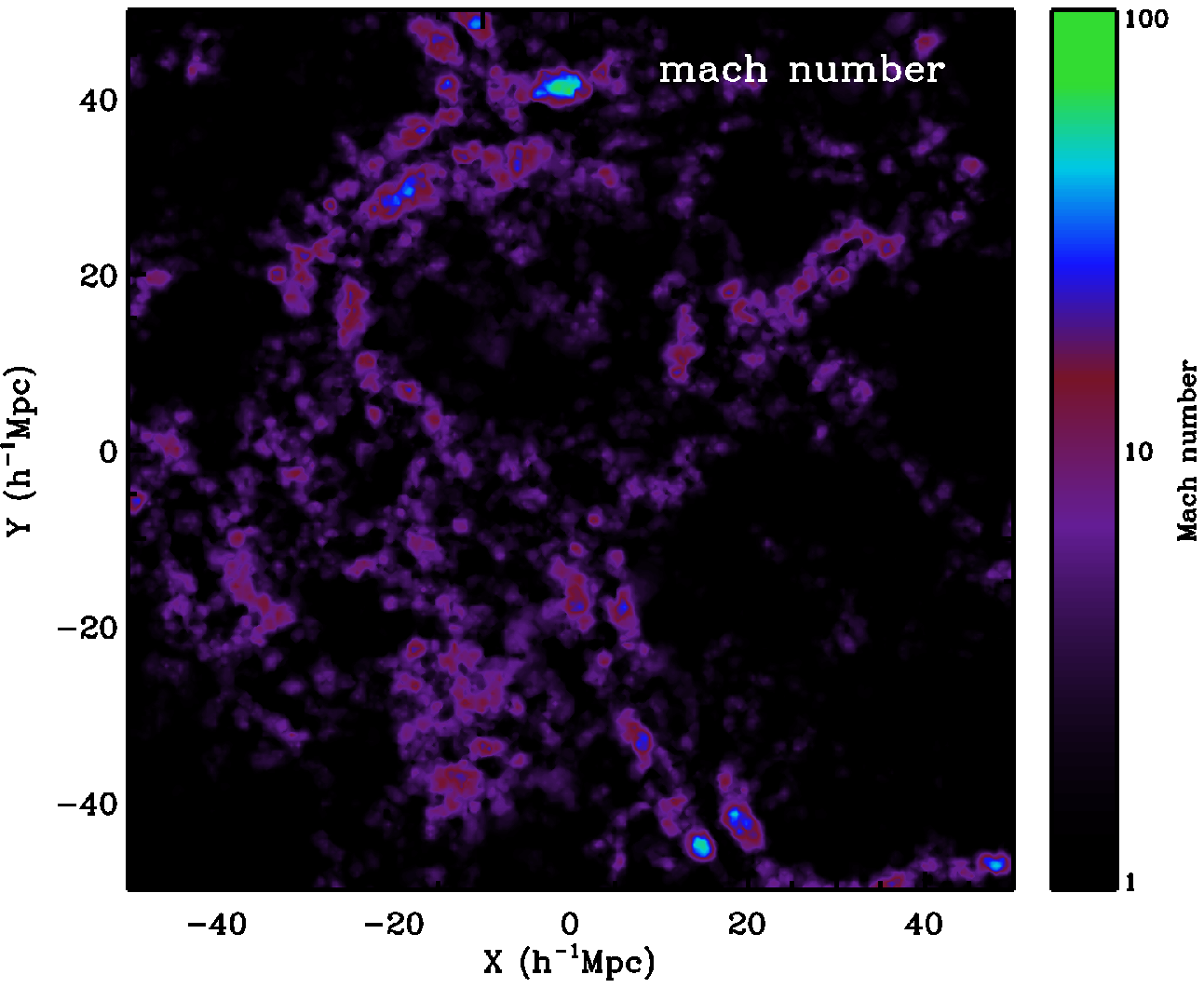}
\caption{\small{Temperature and Mach number in a thin slice of a rich region
of filaments in the MareNostrum simulation (box 2 in 
Fig.~\ref{fig:cosmicweb})}}.
\label{fig:filaments}
\end{figure}

\begin{figure}
\centering
\includegraphics[width=0.43\textwidth]{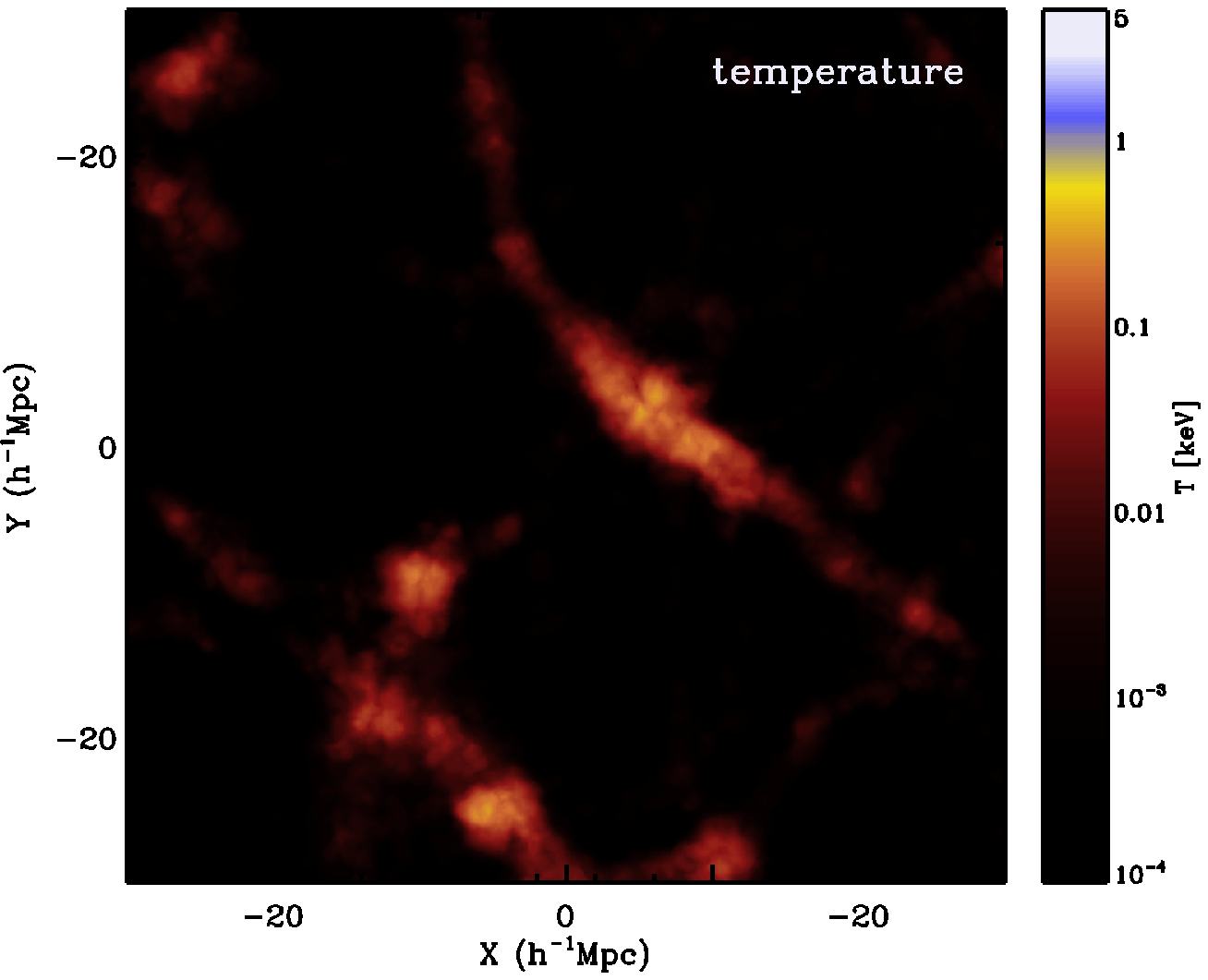}
\includegraphics[width=0.43\textwidth]{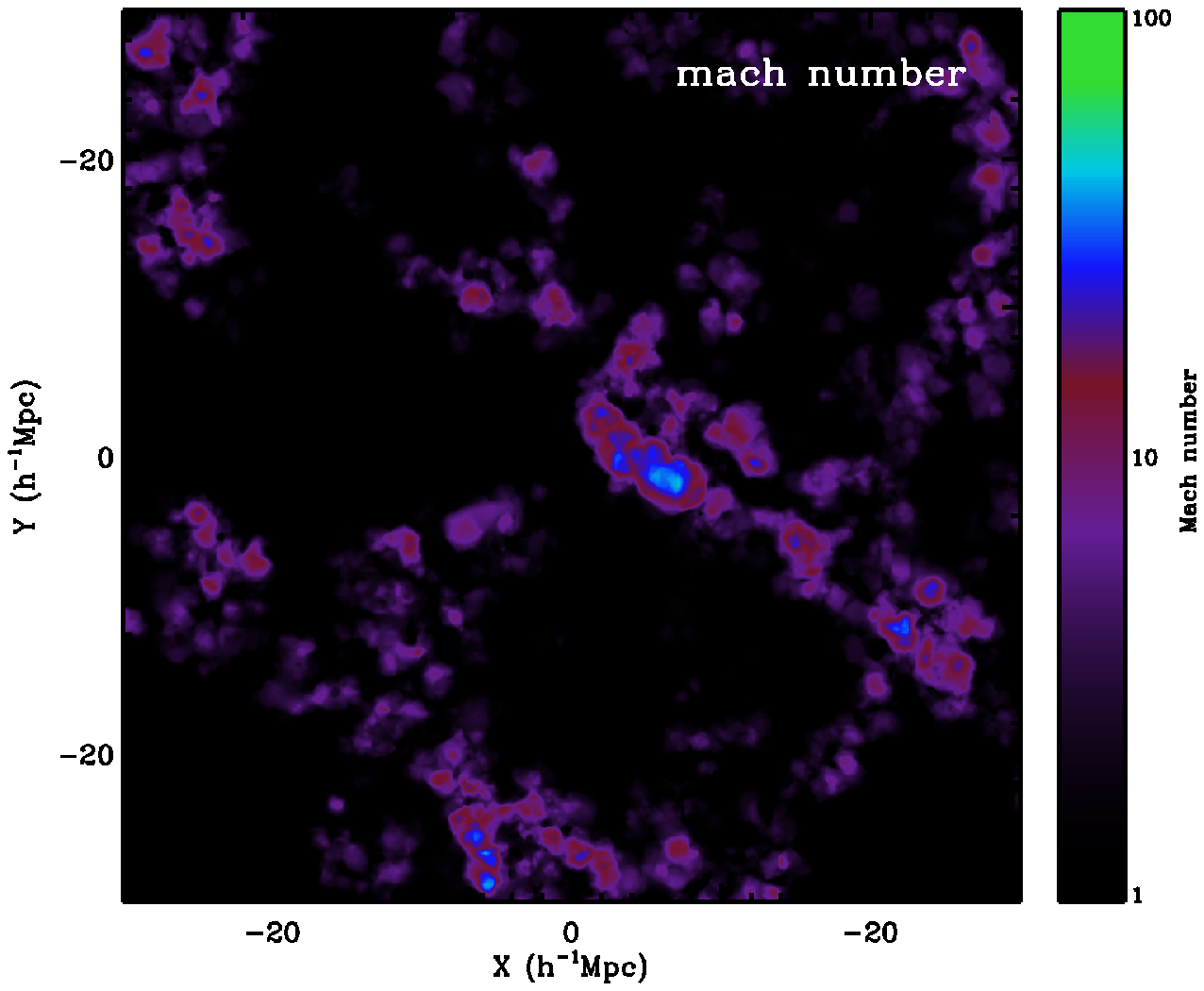}
\caption{\small{Temperature and Mach number in a 10$h^{-1}$Mpc thick slice for 
a void region (box 3 in Fig.~\ref{fig:cosmicweb}).}}
\label{fig:voids}
\end{figure}

Fig.~\ref{fig:voids} depicts a zoom-in of box number 3 of 
Fig.~\ref{fig:cosmicweb}. As in the previous figure, we only consider void 
particles as defined by the SpineWeb procedure. Interesting is to notice that 
voids are not empty. As mentioned before, \cite{gottloeber2003} stated that 
voids are like miniature universes. Voids have what may be called sub-walls and
sub-filaments (a sub-network), with void galaxies the nodes of the 
sub-filaments. These sub-walls and sub-filaments do not have a high enough 
density to be considered walls or filaments on their own. In the region 
chosen, of 60$h^{-1}$Mpc on a side, we see what it could be void galaxies and 
some tenuous filamentary structure. The temperature of this region is 
$T\approx\times10^{3}-10^{6}$. We see that a void region has high Mach numbers
mainly due to the low temperature of the accreting gas. We find that void 
regions in the MareNostrum simulation have shocks with Mach numbers 
$\mathcal{M}\gtrsim 1000$. However, these high Mach
numbers may not be real, and they may be due to the low temperature floor set
in the simulation (see Section~\ref{sec:shockfreq}).
 
\subsection{Radio Emission in the Cosmic Web}
\label{sec:radio}

Each SPH particle has been assigned a radio luminosity via 
Eqn.~\ref{eq:total_emission2} and with the correct setting of 
$f_{\mathrm{A}}(\mathcal{M})$ (Eqn.~\ref{eq:fa}). With this in hand it is
 possible to construct artificial radio maps. To this end, we project the 
emission of each particle long the line of sight and smooth it with the SPH 
kernel size. The radio emission is computed for an observing frequency of 1.4 
GHz and a hypothetical beam of 10$\times$10 arcsec$^{2}$. For comparison, we 
also compute contours of the bolometric X-ray flux. Each SPH particle is 
assigned a luminosity \citep{navarro1995}
\begin{equation}
L_{X}\,=\,1.2\times 10^{-24}\mathrm{erg s^{-1}}\frac{m_{\mathrm{gas}}}{\mu m_{\mathrm{p}}}\frac{n_{e}}{\mathrm{cm^{-3}}}\left(\frac{T}{keV}\right)^{1/2}\,.
\label{eq:xraylum}
\end{equation}

We will study the radio emission in each component of the cosmic web. The
results in these section are obtained using an electron efficiency of
$\xi_{\mathrm{e}}=0.005$. In Section~\ref{sect:radiofunction} we calibrate our 
simulated data against observations, yielding a different electron efficiency
and, therefore, different number of radio objects.

\begin{figure}
\centering
\includegraphics[width=0.39\textwidth,angle=270]{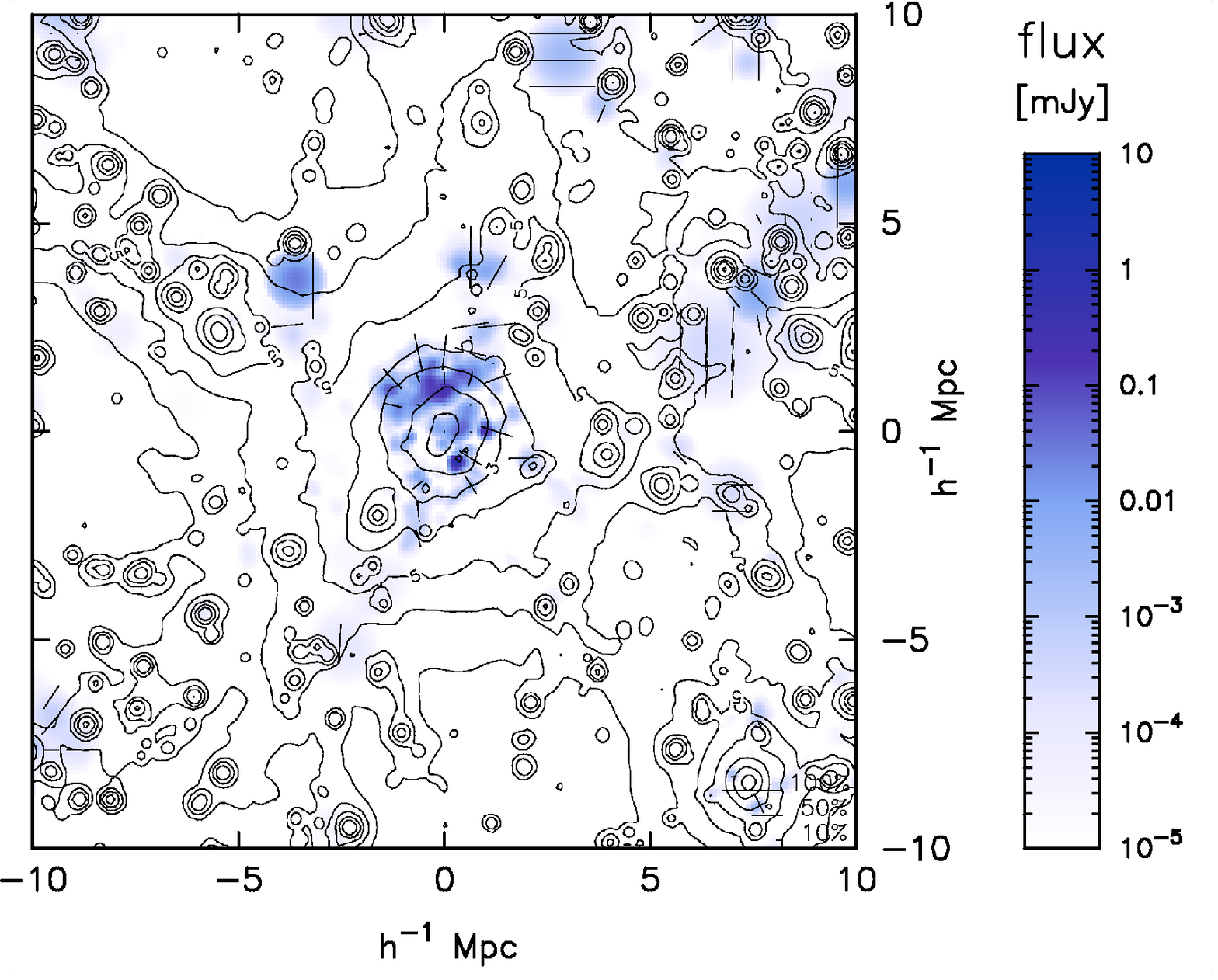}
\includegraphics[width=0.39\textwidth,angle=270]{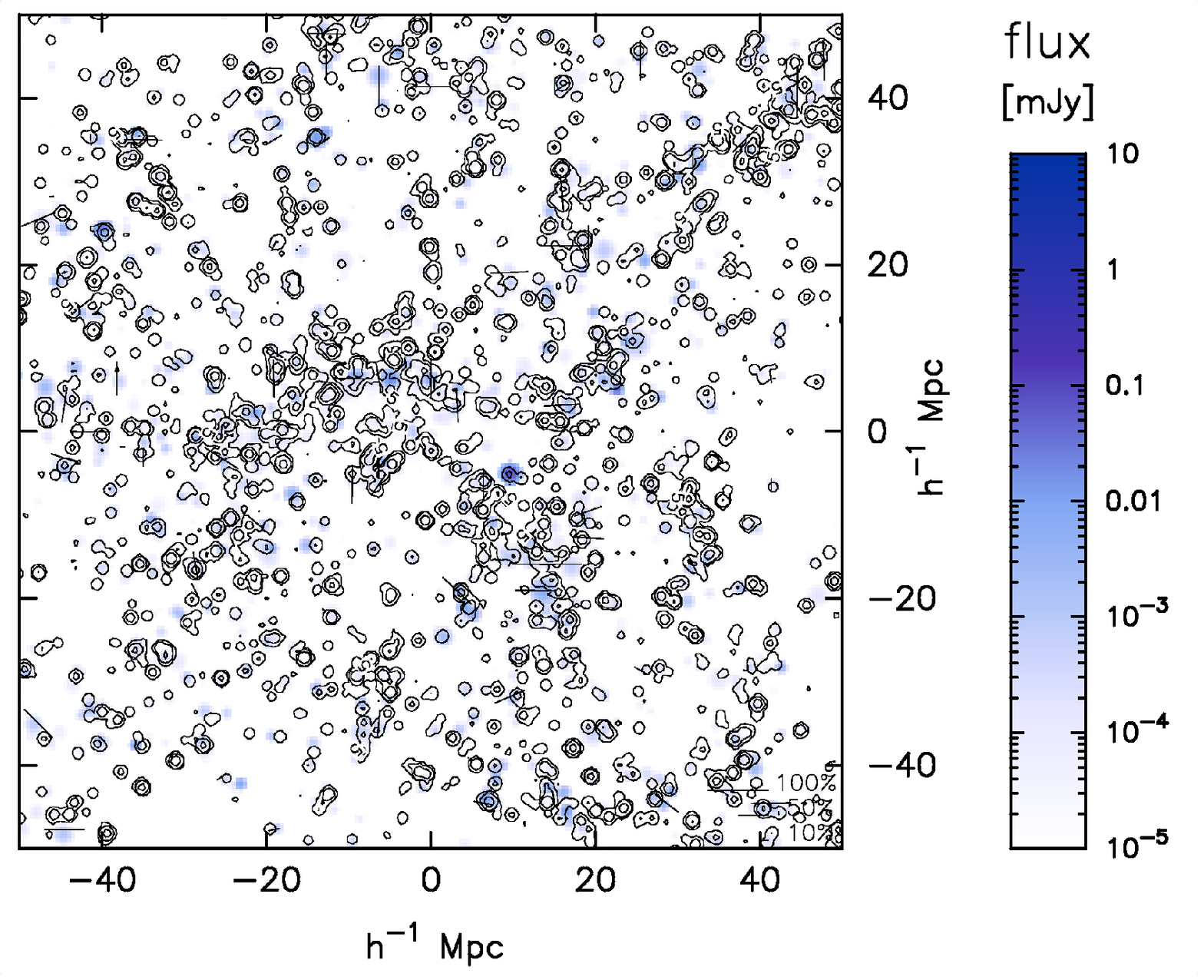}
\includegraphics[width=0.39\textwidth,angle=270]{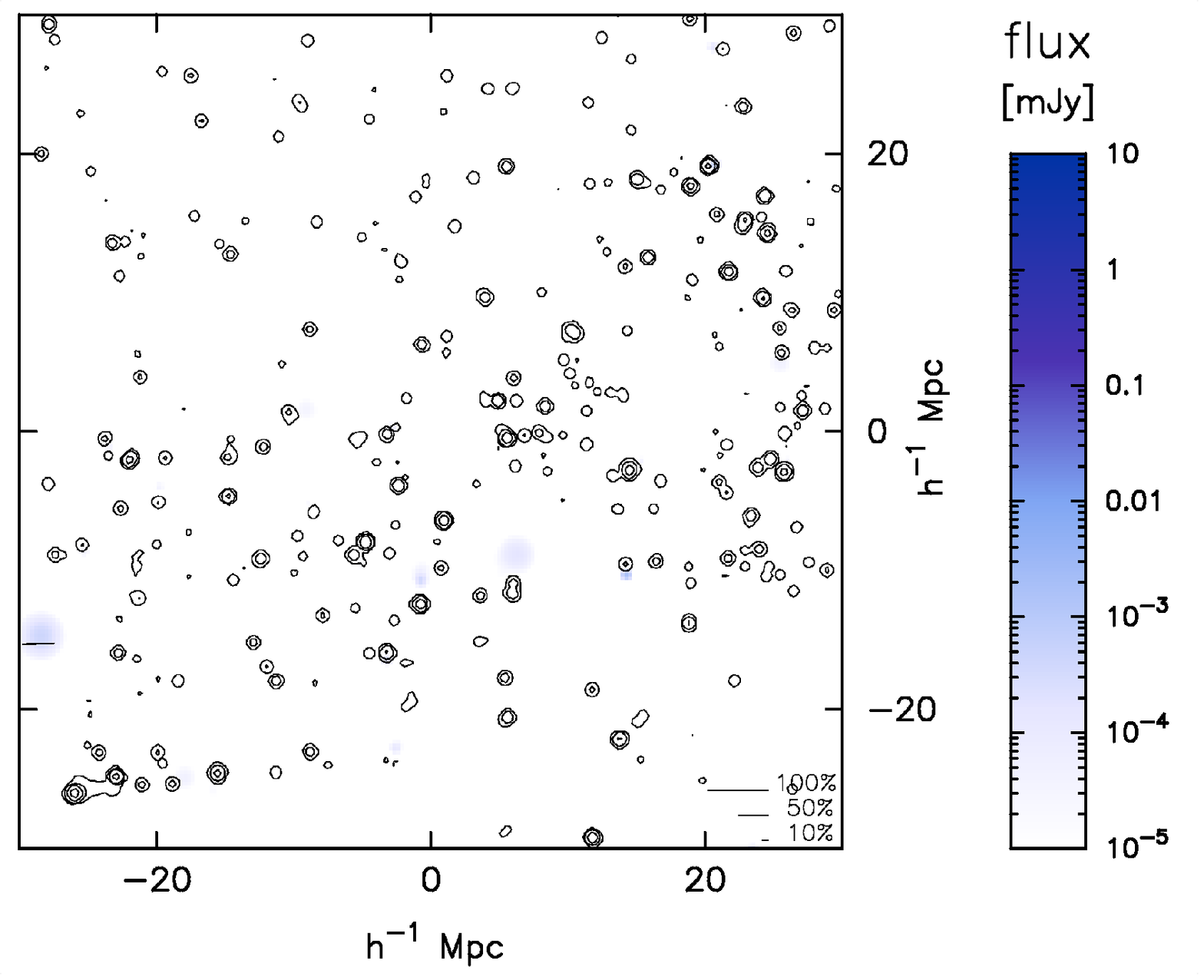}
\caption{\small{Synthetic observations of X-ray and radio emission. The radio
flux at 1.4 Ghz is computed for a hypothetical beam size of $10\times 10$ 
arcsec$^{2}$. Upper panel: the galaxy cluster depicted in 
Fig.~\ref{fig:massivecluster}. Middle panel: the filamentary region shown in
Fig.~\ref{fig:filaments}. Bottom panel: the void region shown in 
Fig.~\ref{fig:voids}. The thin straight lines indicate the direction of 
polarization. The length of the line indicates the degree of polarization.}}
\label{fig:radiolx}
\end{figure}

\subsubsection{Clusters}

Of the total sample of galaxy clusters present in the MareNostrum Universe
($\sim 3800$) we find that only $565$ cluster of galaxies have radio objects 
($693$) with $P_{1.4}>10^{30}$ ergs s$^{-1}$ Hz$^{-1}$ (see 
Sec.~\label{sect:radiofunction} for a correction on this number based on a
correction of $\xi_{\mathrm{e}}$). This represents only $\sim 15\%$ of the total 
galaxy cluster sample. 
%half of that assumed by \cite{ensslin2002} (they assume 
%$1/3$ of clusters have radio halos). 
In total, we find that $\sim 41\%$ of the total sample of galaxy clusters host 
diffuse radio emission. These clusters are not necessarily the most massive 
ones, but they are within the entire mass range.

Fig.~\ref{fig:radiolx} shows the synthetic observation of X-ray and radio
emission of the most massive cluster in the simulation. The thin 
straight lines in the maps indicate the direction of polarization. The length 
of the line indicates the degree of polarization. For comparison the length 
corresponding to 10\%, 50\% and 100\% is shown. The polarization is computed 
according to the formalism described in \citep{burn1966} and 
\citep{ensslin1998}. The radio emission is caused by internal shock fronts in 
the clusters. We could, therefore, exclude external shocks as sources for 
producing radio emission in cluster (see also \citealt{hoeft2008}). We find 
that clusters host the majority of radio objects, with $\sim 20-30$ objects 
with $P_{1.4}>10^{32}$ ergs s$^{-1}$ Hz$^{-1}$ (see 
Section~\ref{sect:radiofunction}). Of a total of $\sim 1500$ radio objects with 
$L_{R}>P_{1.4}>10^{30}$ ergs s$^{-1}$ Hz$^{-1}$  found in the simulation, 
$\sim 500$ are located in galaxy clusters, i.e., 33\% of the total radio 
objects.

We also estimate the relation between the radio luminosity, $P_{1.4}$, of the
radio objects and the emission weighted temperature of its galaxy clusters 
hosts. This result should be similar to the one obtained by \cite{hoeft2008},
only that our sample of galaxy clusters is 10 times larger than theirs.

\begin{figure}
\centering
\includegraphics[width=0.45\textwidth]{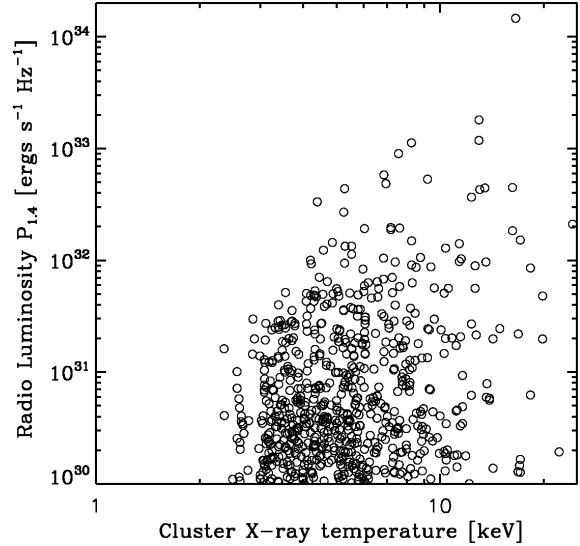}
\caption{\small{Radio luminosity of radio objects versus the emission weighted 
temperature of galaxy clusters.}}
\label{fig:radiopower}
\end{figure}

Fig.~\ref{fig:radiopower} shows that those clusters that host the very
luminous radio objects, i.e., $P_{1.4}>10^{32}$ ergs s$^{-1}$ Hz$^{-1}$, are very
hot ($T_{X}>8$ keV)., i.e., very massive. As was also pointed out by 
\cite{feretti2004}, there is a correlation between the radio luminosity of the
radio objects and the temperature of its host cluster, i.e., the hottest 
clusters hosts the most luminous radio objects.

Our results are quite similar to those presented in \cite{hoeft2008}. This is
expected since their sample is a fraction of the sample of clusters presented
here. The results presented here reinforce that our simple emission model 
recreate diffuse radio emission not only in galaxy clusters, but also in the 
cosmic web, and picks up the trend in the radio luminosity - X-ray relation.

\subsubsection{Filaments}

\cite{bagchi2002} reported the existence of a large-scale diffuse radio 
emission from a large-scale filamentary network. \cite{kim1989} also suggested 
the presence of diffuse radio emission in the Coma Supercluster. Temperatures 
of filaments are in the range $T\approx 10^{5}-10^{7}$, and magnetohydrodynamic 
simulations (MHD) estimate a magnetic field that varies from $10^{-1}$ nG to 
$10$ nG \citep{sigl2003, brueggen2005}. This is in line with 
\cite{ryu2008}, who suggest an average magnetic strength to be of the order
of $10$ nG. However, \cite{bagchi2002} estimated that the strength of the 
magnetic field in the filamentary network is $B\approx 0.3-0.5$ $\mu$G. 
In the same line, \cite{dolag2005} suggests the strength of the 
magnetic field of filaments that connect galaxy clusters to be as large as 
$0.1$ $\mu$G. The middle panel of Fig.~\ref{fig:radiolx} shows the
synthetic observations of the X-ray and the radio emission for the filamentary 
region shown in Fig.~\ref{fig:filaments}. We found that filaments show diffuse 
radio emission, especially in areas near to cluster of galaxies. However, 
although filaments have strong Mach numbers (see Fig.~\ref{fig:filaments}), the
flux in ''pure'' filamentary regions, i.e., regions far from the outskirts of 
clusters is low, as seen in Fig.~\ref{fig:radiolx}, and may not be
detectable with the upcoming radio telescopes. 

As for the number of radio objects, we find that filaments host fewer radio 
objects than clusters, with $\sim 2-3$ radio objects with 
$P_{1.4}>10^{32}$ ergs s$^{-1}$ Hz$^{-1}$ (see Section~\ref{sect:radiofunction}).

\subsubsection{Voids}
 
As discussed in Sect.~\ref{sec:radioemission}, esitmation of magnetic 
field outside galaxy clusters is difficult. Moreoever, MHD simulations have not
been conclusive as well, assigning a magnetic field that varies from $10^{-3}$ 
nG to $10^{-1}$ nG \citep{sigl2003}. Further studies on the magnetic field on 
voids are necessary to precisely determine the strength of these magnetic 
fields.

As can be seen in the bottom panel of Fig.~\ref{fig:radiolx}, there are some
areas in the void region that shows a very tenuous radio emission, with fluxes
$\lesssim 10^{-4}$ mJy. This is basically due to two facts: temperatures in 
voids, as we have seen, is very low compared to the other components of the 
cosmic web. It has also been suggested that the magnetic field in voids is 
quite low. Although observationally it has been possibly to estimate the 
magnetic field in clusters of galaxies \citep[e.g., ][]{bonafede2010}, this has 
not been possible for voids. 
%MHD simulations have not been 
%conclusive as well, assigning a magnetic field that varies from $10^{-3}$ nG to
%$10^{-1}$ nG \citep{sigl2003}. \cite{kim1989} suggest that intergalactic 
%magnetic field may have played an important role in the formation of 
%large-scale structures such as superclusters and voids. Further studies on the 
%magnetic field on voids are necessary to precisely determine the existence of 
%magnetic fields.
%In general, we see that the strongest emission comes from the densest and
%hottest regions of the cosmic web, namely galaxy clusters and large galaxy
%groups \citep[see also][]{skillman2010}. 

\subsection{Radio objects luminosity function}
\label{sect:radiofunction}

As we have shown, there exists the possibility of finding radio emission in
other environments of the cosmic web, such as filaments. We now turn our 
attention to the possibility of finding individual radio objects in the cosmic
web. We do this in a similar way to that of \cite{hoeft2008}, with a slight
difference. We select all particles whose radio emission lies above a very low
threshold. This threshold corresponds roughly to particles with 
$\mathcal{M}\gtrsim 2$. Instead of finding their nearest neighbor and link them
according to their smoothing length, as done by \cite{hoeft2008}, we use HOP to
link the particles and find individual groups. This is a somewhat simpler 
approach, but given that HOP works on a density basis, it assures us that the 
particles will form a compact group. Having found the groups, we compute the
cumulative number of radio objects above a given luminosity, that is, the
luminosity function of diffuse radio objects. 

\begin{figure}
\centering
\includegraphics[width=0.50\textwidth]{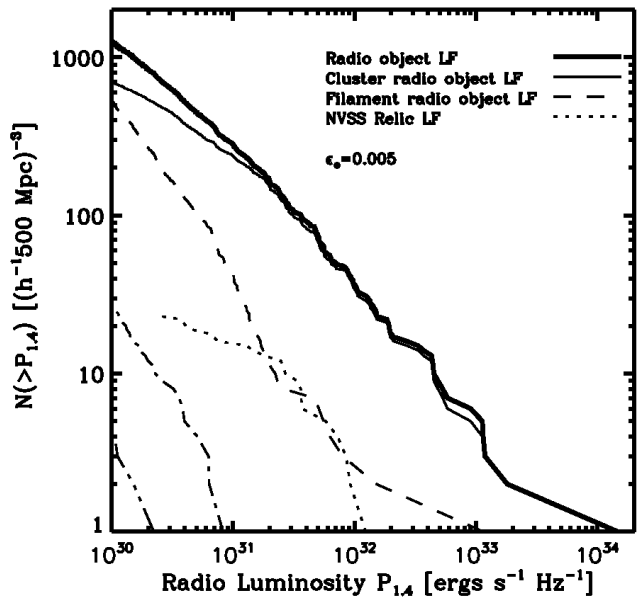}
\caption{\small{Cumulative number density of radio objects. The thick solid 
line indicates the total number of radio objects in the simulation box. The 
thin solid line shows the number of radio objects within clusters. The dashed
line indicates the radio objects present in filaments. The dashed-dotted line
depicts the radio objects in walls, while the dashed-dotted-dashed line
represent the radio objects present in voids. For comparison,
% we depict the
%radio halo luminosity function of En{\ss}lin \& R{\"o}ttgering (2002) (long 
%dashed line) and 
the NVSS relic luminosity function (dotted line).}}
\label{fig:radioemission}
\end{figure}

Fig.~\ref{fig:radioemission} shows the radio luminosity function for radio
objects in the simulation box. For comparison, we also show a radio relic
luminosity function constructed with relic data observations 
\citep{giovannini1999,govoni2001, bonafede2009,clarke2006,feretti2006,
bagchi2006,roettgering1997,vanweeren2009,feretti2001,giacintucci2008,
venturi2007}. Most of these data were obtained from the NRAO VLA Sky Survey 
(NVSS) \citep{condon1998}.
We see that with our simple assumptions, we find a much larger number 
of radio objects in the entire simulation box. Also, if we compare with our 
cluster radio object luminosity function (thin solid line), we still find more 
radio objects. Galaxy clusters host the majority and the most luminous objects,
with $\sim 34$ objects with $P_{1.4}>10^{32}$ ergs s$^{-1}$ Hz$^{-1}$. Filaments 
also host some radio objects, but most of them are less luminous, with only a 
couple with $P_{1.4}>10^{32}$ ergs s$^{-1}$ Hz$^{-1}$. Finally, walls and voids 
also host a few radio objects, but they are rather weak. 
Table~\ref{tab:number_radio_objects} shows the fraction of radio objects by
environment for different total radio luminosity ranges. It can be seen that 
clusters host the majority of more luminous radio objects ($P_{1.4}>10^{32}$ 
ergs s$^{-1}$ Hz$^{-1}$), but when considering the complete sample of radio 
objects, these are located within filaments. This is to expect, since a great
part of the mass of the Universe resides in filaments \citep{aragoncalvo2010b}.
This is also an indication that the detection of diffuse radio emission in
filaments could shed light on the missing-baryon problem. Walls also host a 
great number of low-luminous radio objects, while in voids the presence of 
radio objects is negligible.

\begin{figure}
\centering
\includegraphics[width=0.50\textwidth]{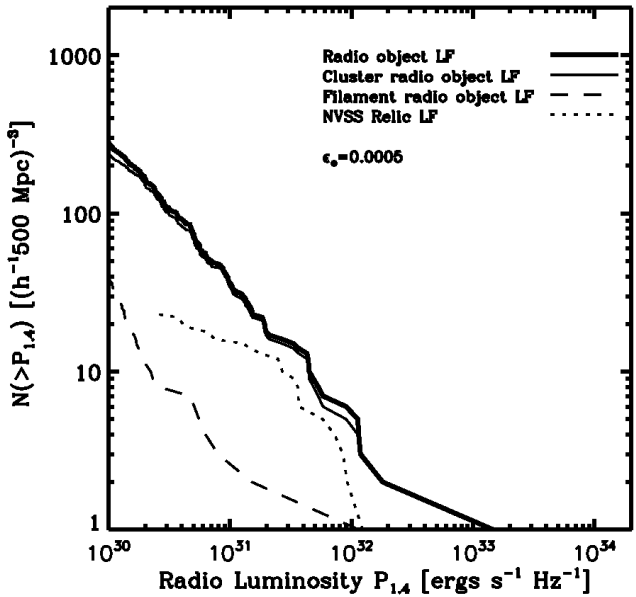}
\caption{\small{Same as Fig.~\ref{fig:radioemission}, but the total radio power
has been rescaled by using a electron efficiency parameter of 
$\epsilon_{e}=0.0005$. By doing this, radio objects are less luminous, and we
are able to match our simulated relics with observational NVSS relics.}}
\label{fig:radio_lum_scaled}
\end{figure}

As pointed out by \cite{hoeft2008}, the overestimation of radio objects may be
due to an overestimation of the electron efficiency ($\xi_{\mathrm{e}}$), as well
as the magnetic fields. The value of the the electron efficiency parameter is 
quite uncertain from current observational and theoretical constrains
\citep{skillman2010}. From Eqn.~\ref{eq:total_emission2}, we see that the 
relationship between the total radio power and $\xi_{\mathrm{e}}$ is linear, so 
in order to obtain less luminous objects, we just have to rescaled the total 
radio power, e.g., an electron efficiency of $\xi_{\mathrm{e}}=0.0005$ will lead 
to a total radio power that is low by a factor of 10. In order to match our 
results to observations, i.e., the NVSS radio relic luminosity function,
%the radio halo luminosity function of 
%\cite{ensslin2002}, 
we rescale the luminosity function using $\epsilon_{e}=0.0005$, shown in 
Fig.~\ref{fig:radio_lum_scaled}. As expected, radio objects are less luminous, 
and therefore, less radio objects are found. In the plotted range, there are no
radio objects in voids and filaments, only 1 in filaments, and some group of 
ten in clusters. With this electron efficiency, the cluster radio objects 
matches the observational NVSS radio relic luminosity function 
(see Fig.~\ref{fig:radio_lum_scaled}). However, we predict more relics at the
high and low power compared with the observed function. This can be to due to 
the fact that the observed relics, to date, is still small, compared to the 
large number of relics found in the simulation with our method. Upcoming radio
observations will increase the number of observed relics, making the statistics
stronger. Table~\ref{tab:number_radio_objects} shows the fraction of radio 
objects by environment for different total radio luminosity ranges with the 
adjusted electron efficiency parameter. A study of the effects of the magnetic 
fields is beyond the scope of this paper, so we do not investigate this.

The luminosity function (Eqn.~\ref{fig:radioemission}) suggest what can be 
expected with the use of the upcoming radio telescopes. An increase of the 
surface brightness sensitivity by a factor of 10 will result in an increase of
the number of radio objects by a factor of 10-100 (see also \cite{skillman2010}.
We can also estimate the number of radio objects to be found in a given survey 
area and redshift depth by solving the equation:
\begin{equation}
\frac{dV}{dzd\Omega}(z)\,=\,D_{H}\frac{(1+z)^{2}D^{2}_{A}}{E(z)}\,,
\label{eq:covol1}
\end{equation}
where $D_{H}\equiv c/H_{0}$ is the \emph{Hubble distance}, $D_{a}$ is the 
angular diameter distance and 
$E(z)=\sqrt{\Omega_{m}(1+z)^{3}+\Omega_{k}(1+z)^{2}+\Omega_{\Lambda}}$. For a flat,
$\Lambda$-dominated Universe like the MareNostrum simulation, $\Omega_{k}=0$ 
and Eqn.~\ref{eq:covol1} can be rewritten as \citep{carroll1992}
\begin{equation}
V_{C}\,=\,\frac{4\pi}{3}D_{M}^{3}\,,
\label{eq:covol2}
\end{equation}
where $D_{M}$ is the comoving distance, which, for a flat Universe, coincides 
with the line-of-sight comoving distance. Eqn.~\ref{eq:covol2} corresponds to 
the total comoving volume in an all-sky survey out to redshift $z$. For the 
MareNostrum Universe, the result of Eqn.~\ref{eq:covol1} corresponds to an 
all-sky survey out to $z\sim 0.18$. If we assume an electron efficiency of 
$\xi_{\mathrm{e}}=0.005$ ($\xi_{\mathrm{e}}=0.0005$), we expect to find 34 (4) 
radio objects in 30 (3) galaxy clusters, 2 (1) radio objects in filaments and 
none in walls and voids, all with a total radio luminosity of $P_{1.4}>10^{32}$ 
ergs s$^{-1}$ Hz$^{-1}$  within this cosmological volume. Assuming that the 
number density of radio objects is nearly constant through cosmic time 
\citep{skillman2010}, we could estimate that in an all-sky survey out to 
$z=0.5$ (which corresponds to a volume of $\sim15$ times the previous one), we 
could find 510 (60) radio objects in 450 (45) clusters, 30 (15) radio 
objects in filaments and a few in walls and voids.

\begin{table}
 \caption{Fraction of total radio objects ($\sim 14600$) with a given radio 
luminosity $P_{1.4}$ in terms of their environment assuming an electron 
efficiency of $\xi_{\mathrm{e}}=0.005$ ($\xi_{\mathrm{e}}=0.0005$) Units of radio 
luminosity given in ergs s$^{-1}$ Hz$^{-1}$.}
% \begin{minipage}{0.80\linewidth}
   \centering
   \begin {tabular}{|l||c|c|c|}
     \hline
     & & \\
     & $P_{1.4}>10^{32}$ &  $10^{30}<P_{1.4}\le10^{32}$  & $P_{1.4}<10^{30}$ \\
     & & \\
     \hline
     \hline
     & & \\
     Clusters & 0.23\% (0.03\%)  & 4.51\%  (1.6\%) &  10.33\% (13.46\%) \\
     Filaments& 0.01\% (0\%)       & 3.65\%  (0.3\%) &  74.24\% (77.58\%) \\
     Walls    & 0\%    (0\%)    & 0.18\%  (0\%)   &   6.40\%  (6.58\%) \\
     Voids    & 0\%    (0\%)    & 0.03\% (0\%)   &   0.42\%  (0.45\%) \\
%     Clusters & 34 & 659 &  1510 \\
%     Filaments& 2  & 533 & 10855 \\
%     Walls    & 0  &  27 &   935 \\
%     Voids    & 0  &   4 &    62 \\
     & & \\
     \hline
   \end {tabular}
% \end{minipage}
 \label{tab:number_radio_objects}
\end {table}

The use of the SpineWeb enable us to make predictions of the radio flux in
filaments. Assuming an spectral index of $\alpha=1$, filaments at a redshift
$z\sim 0.15$, in a frequency of 150 MHz, a radio flux of $S_{150 MHz}\sim 0.12$
$\mu$Jy.

\section{Conclusions}
\label{sec:conclusions}

We have analyzed one of the largest hydrodynamical simulations of cosmic 
structure formation, namely the MareNostrum simulation, with the purpose of
studying the possibility of detecting radio emission in the cosmic web. 
To that purpose, we use: i)  a novel method for estimating the radio emission of
strong shocks that occur during the process of structure formation in the 
Universe developed by \cite{hoeft2007} and ii) the SpineWeb technique 
The SpineWeb procedure deals with the topology of the underlying density field,
correctly identifying voids, walls and filaments. While previous studies 
focused on the study of cosmological shocks in selected catalogs of galaxy 
clusters or by defining the different environments of the cosmic web by density
and/or temperature cuts, we are able to correctly disentangle the complex 
filamentary network of the cosmic web. The computation of the radio emission is
based on an estimate for the shock surface area. The radio emission is computed
per surface area element using the Mach number of the shock and the downstream 
plasma properties.

In order to properly calculate the radio emission, two corrections were 
made.  The first correction deals with the shock surface area. 
The constant area factor ($f_{a}$) of Eqn.~\ref{eq:total_emission2} was 
corrected using shock tube tests and a convolution method described in
Appendix~\ref{app:dsdm}. 
One deals with the temperature of the simulation. Given that the 
simulation was run with a low floor temperature and without UV background, 
expansion cooling leads to very low temperature, which leads to very high 
Mach numbers. We corrected this by scaling the temperature of cold, low density
regions using $T=T_{0}(1+\delta)^{\gamma-1}$ \citep{hui1997}, where $\delta$ is 
the overdensity.
The first step was to calculate the strength of cosmological shocks, 
characterized by their Mach number using the method described in
 \cite{hoeft2008}. We find that each environment of the cosmic web has a 
distinct Mach distribution: voids have a characteristic Mach number of 
$\mathcal{M}_{\mathrm{voids}}\approx 18$; walls, 
$\mathcal{M}_{\mathrm{walls}}\approx 7.5$; filaments, 
$\mathcal{M}_{\mathrm{filaments}}\approx 6.2$; and clusters, 
$\mathcal{M}_{\mathrm{clusters}}\approx 1.8$. The strength of the shock, i.e., the
characteristic Mach number, is closely related to the temperature and density
of the medium. Voids are the coldest and less dense region in the 
Universe, with a mean temperature 
$\langle\mathrm{T}\rangle_{\mathrm{voids}}\approx 7\times 10^{5}$ K and 
$\langle\delta\rangle_{\mathrm{voids}}\approx 2.9$, and they interact with walls 
and filaments, which are denser and hotter structures, with filaments reaching 
a mean temperature of 
$\langle\mathrm{T}\rangle_{\mathrm{voids}}\approx 7.3\times 10^{6}$ K and 
$\langle\delta\rangle_{\mathrm{voids}}\approx 72$.

Applying a radio emission model based on DSA to the simulation we are able to
reproduce the diffuse radio emission in galaxy clusters using a lower 
electron efficiency to that used by \cite{hoeft2008}. This diffuse radio 
emission is also present in filaments. In walls and voids the emission is very 
weak, resulting in almost no presence of radio objects. By identifying radio 
objects based on their density, we find that clusters hosts the majority of 
such objects. Assuming an electron efficiency of $\xi_{\mathrm{e}}=0.005$, we 
find that 30 galaxy clusters (out of 3865) host 34 radio objects with 
$P_{1.4}>10^{32}$ ergs s$^{-1}$ Hz$^{-1}$. If we consider a radio luminosity
of $10^{30}<P_{1.4}\le10^{32}$ ergs s$^{-1}$ Hz$^{-1}$, then 659 radio objects
are found in 547 galaxy clusters, only $\sim 15\%$ of the total sample. These 
radio objects are present in the entire mass range of the galaxy clusters, 
i.e., both low mass and high mass clusters present radio objects. Filaments 
also host radio objects, although only 2 of them are very luminous 
($P_{1.4}>10^{32}$ ergs s$^{-1}$ Hz$^{-1}$). However, the majority of low luminous
radio objects ($P_{1.4}<10^{30}$) are present in filaments. The presence of 
radio objects in walls and voids is rather negligible. 

We calibrated our simulated data against the observational NVSS radio relic
luminosity function, yielding an electron efficiency of 
$\xi_{\mathrm{e}}=0.0005$. This is a factor of 10 less than the one estimated by
\cite{hoeft2008}. When using this efficiency, the number of very luminous radio
objects drop dramatically. We only find 4 radio object with $P_{1.4}>10^{32}$ 
ergs s$^{-1}$ Hz$^{-1}$ and 1 in filaments. In the range 
$10^{30}<P_{1.4}\le10^{32}$ ergs s$^{-1}$ Hz$^{-1}$, we find 235 radio objects in
198 galaxy clusters (5\% of the total sample) and 43 in filaments.

An all-sky survey up to $z=0.5$ and assuming an efficiency of 
$\xi_{\mathrm{e}}=0.005$ ( $\xi_{\mathrm{e}}=0.0005$) should result in the 
discovery of 510 (60) radio objects in 450 (45) clusters, and 30 (15) radio 
objects in filaments. With the increase of surface brightness sensitivity of 
the upcoming radio telescopes, the detection of radio objects should increase 
by a factor of 10, which opens in the possibility of finding a considerable 
number of radio objects in filaments. Furthermore, we predict that the radio 
flux of filaments at redshift $z\sim 0.15$, and at a frequency of 150 MHz, 
should be $S_{\small{150 MHz}}\sim 0.12$ $\mu$Jy.

\section*{Acknowledgments}

The authors would like to thank Gustavo Yepes for making the MareNostrum 
simulation available. PAAM gratefully acknowledges Bernard Jones for helpful 
discussions and support by the German Federal Ministry for Education and 
Research under the {\it Verbundforschung}.

\appendix

\section{Correcting the shock surface area distribution}
\label{app:dsdm}

Convolution is a mathematical operation on two functions, $f$ and $g$, which
produces a third function that is a modified version of the one of the original
functions. Usually, one of the two functions is taken to be a \emph{kernel} 
function which acts on the other function, modifying it. The standard 
convolution states that
\begin{equation}
h(t)\,=\,f(t)\ast g(t)\,\equiv\,\int_{-\infty}^{\infty}f(\tau)g(t-\tau)d\tau\,,
\label{eq:normal_convolution}
\end{equation}
where $\ast$ is the convolution operand. The inverse of this operation is 
called deconvolution. In our case, we have a measured distribution, namely 
$dS/d\log{\mathcal{M}}$, obtained from the simulation and we wish to obtain the
\emph{real} distribution, i.e., the shock surface area distribution taking into
account the particles that were assign a low Mach number in strong shocks and 
the underestimation of the different Mach numbers. Therefore, we need to find
the shock tube \emph{distribution kernel} taken from the shock tubes data in 
order to deconvolve it with our measure function. However, instead of finding 
the inverse of the kernel function, we will iterate the convolution procedure, 
which will yield the real distribution. The convolution we want to solve is
\begin{equation}
q^{m}(\log{\mathcal{M}})\,=\,\int_{0}^{\infty}d\log{\mathcal{M}}'
f(\log{\mathcal{M}},\log{\mathcal{M}}')q^{r}(\log{\mathcal{M}}')\,,
\label{eq:our_convolution}
\end{equation}
where $q^{m}(\log{\mathcal{M}})$ is the \emph{measured} shock surface 
distribution, $f(\log{\mathcal{M}},\log{\mathcal{M}}')$ is the kernel function
and $q^{r}(\log{\mathcal{M}})$ is the \emph{real} distribution we
want to find. In Eqn.~\ref{eq:our_convolution} we have taken two things into 
account: i) there are not negative Mach numbers, and ii) the kernel function 
depends on the Mach number.

In Section~\ref{sec:shocktubes} we saw that a simple representation to the 
shock surface area of the shock tubes are a Heaviside function and a Gaussian. 
Therefore, our kernel has the form
\begin{equation}
\mathbf{F}(\log{\mathcal{M}})_{ij}\,=\,\mathcal{K}\cdot\mathcal{H}
(\log{\mathcal{M}}_{i})+\mathcal{G}_{ij}(\log{\mathcal{M}}_{i},
\log{\mathcal{M}}_{j})\,,
\label{eq:kernel}
\end{equation}
where $\mathcal{G}_{ij}$ is a Gaussian with form
\begin{equation}
\mathcal{G}_{ij}\,=\,e^{-(\log{\mathcal{M}}-
\log{\mathcal{M}'})^{2}/2\sigma(\log{\mathcal{M}'})^{2}}\,,
%\mathcal{G}_{ij}\,=\,\frac{1}{2\pi\sigma^{2}}e^{-(\log{\mathcal{M}}-
%\log{\mathcal{M}'})^{2}/2\sigma(\log{\mathcal{M}'})^{2}}\,,
\label{eq:gaussian}
\end{equation}
where the standard deviation also depends on the measured Mach numbers, i.e.,
$\sigma=\sigma(\log{\mathcal{M}})$ as seen in the previous section.

After constructing the kernel with the fitting parameters of the various shock
tubes, we give a guess function in order to start the iteration procedure. As
a first guess, we choose $q^{r}_{1}=q^{m}$, and convolve that with the kernel,
giving as a result $q^{m}_{1}$. The new guess is then calculated by subtracting
the previous guess with the difference between the result of the convolution
of the first guess with the measured function, i.e., 
$q^{r}_{2}=q^{r}_{1}-(q^{m}_{1}-q^{m})$. We found that after iterating 5 times we
are able to recover $q^{m}$.

\bsp

\label{lastpage}

\end{document}